\documentclass{pasj00}
\draft

\begin{document}
\SetRunningHead{Author(s) in page-head}{Running Head}
%\Received{}%{yyyy/mm/dd}
%\Accepted{}%{yyyy/mm/dd}
%\Published{}%{yyyy/mm/dd}

\title{The New Primary X-ray Component Confirmed in the Seyfert I Galaxy IC4329A}

%%% begin:list of authors
% Do NOT capitalize all letters in "textsc".
\author{Katsuma~\textsc{Miyake}\altaffilmark{1}, 
Hirofumi~\textsc{Noda}\altaffilmark{2}, 
Shin'ya~{\sc Yamada}\altaffilmark{3}, 
Kazuo~{\sc Makishima}\altaffilmark{4}, and
Kazuhiro~{\sc Nakazawa}\altaffilmark{1}}
%\affil{University of Tokyo, 7-3-1 Bunkyo, Tokyo, Japan}\email{ccccc@xxx.xxx.xx.xx}
\altaffiltext{1}{University of Tokyo, 7-3-1 Bunkyo, Tokyo, 113-0033, Japan}
\altaffiltext{2}{High Energy Astrophysics Laboratory, Nishina Center, RIKEN, 2-1 Hirosawa, Wako city, Saitama 351-0198, Japan}
\altaffiltext{3}{Tokyo Metropolitan University, 1-1 Minami Osawa, Hachioji city, Tokyo, Japan}
\altaffiltext{4}{MAXI team, Global Research Cluster, RIKEN, 2-1 Hirosawa, Wako city, Saitama 351-0198, Japan}

%%% Please use the following style in case that sorting by 
%%% affiliation is impossible. 
% v
% \author{%
%   D-Firstname \textsc{D-Familyname}\altaffilmark{1}
%   E-Firstname \textsc{E-Familyname}\altaffilmark{1,2}
%   and
%   F-Firstname \textsc{F-Familyname}\altaffilmark{2}}
% \altaffiltext{1}{Address of Institute}
% \email{ddddd@xxx.xxx.xx.xx}
% \email{eeeee@xxx.xxx.xx.xx}
% \altaffiltext{2}{Address of Institute}

%% `\KeyWords{}' always has to be placed before `\maketitle'.
\KeyWords{galaxies: active --- galaxies: individual (IC4329A) --- galaxies: Seyfert --- X-rays: galaxies} %Do NOT move this preamble from here!

\maketitle

\begin{abstract}
The bright and highly-variable Seyfert I Active Galactic Nucleus, IC4329A, was observed with Suzaku 5 times in 2007 August with intervals of $\sim 5$ days, and a net exposure of 24--31~ks each. Another longer observation was carried out in 2012 August with a net exposure of 118~ks. In the 6 observations, the source was detected in 2--45~keV with average 2--10~keV fluxes of $($0.67--1.2$)\times10^{-10}$~erg~cm$^{-2}$~s$^{-1}$. The intensity changed by a factor of 2 among the 5 observations in 2007, and 1.5 within the 2012 observation. 
Difference spectra among these observations revealed that the variability of IC4329A was carried mainly by a power-law component with a photon index $\Gamma \sim 2.0$. However, in addition to this primary component and the associated reflection, the broad-band Suzaku data required another, harder, and less variable component with $\Gamma \sim 1.4$. The presence of this new continuum was also confirmed by analyzing the same 6 data sets through the spectral decomposition technique developed by \citet{Noda2013a}. This $\Gamma \sim 1.4$ continuum is considered to be a new primary component that has not been recognized in the spectra of IC4329A so far, although it was recently identified in those of several other Seyfert I galaxies (Noda et al. \yearcite{Noda2013a}, \yearcite{Noda2014}).
\end{abstract}

\section{Introduction}\label{sec:intro}

So far, X-ray spectra of Seyfert I Active Galactic Nuclei (AGNs) have been considered to consist of a single power-law (PL) shaped primary emission, and its reflection component accompanied by an Fe K$\alpha$ line (\cite{FabianMiniutti2005}). The PL component, with a typical photon index of $\Gamma\sim2.0$, is thought to be generated by inverse Compton scattering in a ``corona" near a Super Massive Black Hole (SMBH) in the AGN (e.g., \cite{Haardt1994}). By materials surrounding the SMBH, a fraction of these primary X-ray photons are reprocessed to form a reflection continuum with Fe-K absorption edge and an Fe-K line, via combined effects of photo-absorption and Compton scattering (\cite{GeorgeFabian1991}). Some of this secondary emission that is generated in a region near the SMBH is considered to exhibit a broad Fe K$\alpha$ line, extending from $\sim \ 4$~keV to $\sim \ 7$~keV due to relativistic effects. Apart from these secondary signals, a fraction of the primary photons are thought to be absorbed by these materials, to form so-called partially absorbed continuum (e.g., \cite{Holt1980}).\par
Although the above spectral composition has been brought into a general consensus, unambiguous separation of the primary continuum from the secondary reflection components without any assumptions is in reality difficult, because they are both relatively featureless except for the Fe-K line and edge structures. Thus, an X-ray spectrum of an AGN can be interpreted by various models (e.g., \cite{Cerruti2011}), often with different astrophysical implications. Furthermore, the widely employed assumption, that the primary emission from an AGN is represented by a single PL continuum, has not been confirmed observationally. Then, timing information is considered useful in arriving at less ambiguous spectral decompositions, because different components are expected to show different variability. Along the above consideration, \citet{Churazov2001} used Count-Count Plots (CCPs), and decomposed an X-ray spectrum of Cygnus X-1 into a variable and a stationary components. \citet{Taylor2003} applied the same method to spectra of some Seyfert AGNs to achieve similar results. \par
Noda et al. (\yearcite{Noda2011a}, \yearcite{Noda2011b}) systematically performed these attempts on broad-band X-ray spectra of AGNs, and developed the ``Count-Count Correlation with Positive Offset" (C3PO) method (\cite{Noda2013b}). By this model-independent technique, they revealed for the first time that X-ray spectra of many AGNs harbor two distinct primary components (Noda et al. \yearcite{Noda2011a}, \yearcite{Noda2013a}, \yearcite{Noda2014}); one is a highly variable and steeper PL with $\Gamma\sim2.0$ without strong absorption, while the other is a flatter $\Gamma\sim1.4$ PL, with  slower variation and stronger absorption. The former becomes dominant when the luminosity increases typically above $\sim0.01$ times the Eddington limit. The latter, indeed a new discovery, is thought to have been regarded previously as a partially absorbed part of the former component, even though we now know that the two PL continua differ both in their spectral slopes and variability characteristics. \par
In this way, the C3PO method has succeeded in significantly eliminating the spectral modeling degeneracy in an assumption-free manner. However, its outcome is so innovative that we need to reinforce the method by more conventional analyses. One of them, to be adopted in the present paper, is so-called difference spectrum method, which studies spectral differences when the source is brighter and fainter. For this purpose, we need to select radio-quiet Seyfert I AGNs, in order for the X-ray signals not to be contaminated by possible jet emission. In addition, the objects must have high variability and sufficient X-ray flux, and be observed for a long time compared to their variability time scales.\par 
As an example satisfying the above requirements, we selected IC4329A, which was observed with Suzaku 6 times. This is a highly variable (\cite{Perola1999}) Seyfert 1 AGN, with a redshift of $z=0.016$ (\cite{Willmer1991}) and an estimated mass of $1.3 \times 10^8M_{\odot}$ (\cite{Markowitz2009}). It shows a broad-band continuum with $\Gamma\sim2$ (e.g. \cite{Piro1990}), together with a narrow Fe K$\alpha$ line at $\sim6.4$~keV and an absorption edge at $\sim7.1$~keV (\cite{Gondoin2001}). These features are typical of type I Seyferts. With a recent NuSTAR observation, a broad Fe-K line with a width of $\sigma=0.33$~keV was detected at a rest-frame energy of 6.46~keV, and the cut-off energy of the hard X-ray spectrum was estimated at $\sim186$~keV (\cite{Brenneman2014}). With these in mind, we analyze the 6 Suzaku data sets of IC4329A for quantification of the broad band continua. Unless otherwise stated, the errors in the present paper refer to 90\% confidence limits. \par
%This AGN was observed by Suzaku 5 times in 2007 and once in 2012. 
%As shown in figure \ref{fig:lc}, we observed clear variability in soft X-rays, around a mean 2-10~keV flux of $7.0 \times 10 ^{-11}$~erg~cm$^{-2}$~s$^{-1}$.

\section{Observation and data reduction}
As already mentioned, IC4329A was observed with Suzaku 5 times in 2007 August, and once in 2012 August. Table \ref{tab:obs} shows the log of these observations, together with their abbreviations. We analyze public data of XIS 0, XIS 3, and HXD-PIN obtained on these occasions, wherein the XIS and the HXD were both operated in their normal modes. \par
In analyzing the XIS data, we co-added the events from XIS 0 and XIS 3. Their on-source events were extracted from a circular region of $3'$ radius, and background events from an annular region of radii $4'.8 - 7'.8$. The response matrices and ancillary response files were created by {\tt xisrmfgen} and {\tt xissimarfgen} (\cite{Ishisaki2007}), respectively. From the HXD-PIN data, we subtracted Non X-ray Background (NXB) created by the NXB model (\cite{Fukazawa2009}), and the Cosmic X-ray Background employing the spectral form of \citet{Boldt1987}. Systematic uncertainties in this background subtraction process are estimated as 3.8\%, which is appropriate for a data accumulation of 3~ks or longer \citep{Fukazawa2009}.

\begin{table}
  \caption{The log of Suzaku observations of IC4329A.}\label{tab:obs}
  \begin{center}
    \begin{tabular}{lllllll}
      \hline
      ID & start date && end date && exposure$^*$ & flux$^\dagger$ \\ \hline
      Obs.07a & 2007 Aug. 01 & 05:28:20 & Aug. 02 & 01:00:24 & (25~ks, 21~ks) & $1.0$ \\
      Obs.07b & 2007 Aug. 06 & 00:41:45 & Aug. 06 & 21:41:45 & (31~ks, 26~ks) & $1.2$ \\
      Obs.07c & 2007 Aug. 11 & 11:00:23 & Aug. 12 & 05:06:24 & (27~ks, 24~ks) & $1.1$ \\
      Obs.07d & 2007 Aug. 16 & 11:48:47 & Aug. 17 & 03:52:14 & (24~ks, 20~ks) & $1.0$ \\
      Obs.07e & 2007 Aug. 20 & 23:26:16 & Aug. 21 & 13:20:24 & (24~ks, 19~ks) & $0.67$ \\
      Obs.12 & 2012 Aug. 13 & 02:13:09 & Aug. 15 & 15:25:17 & (118~ks, 113~ks) & $1.0$ \\
      \hline
    \end{tabular}
  \end{center}
  \begin{itemize}
    \item[$^*$] (XIS, HXD). 
    \item[$^\dagger$] In 2--10~keV, in units of $10^{-10}$~erg~cm$^{-2}$~s$^{-1}$.
  \end{itemize}
\end{table}

\section{Data analysis}

\subsection{Light curves and spectra}
\begin{figure}[h!]
 \begin{center}%\centering
  \includegraphics[width=16cm]{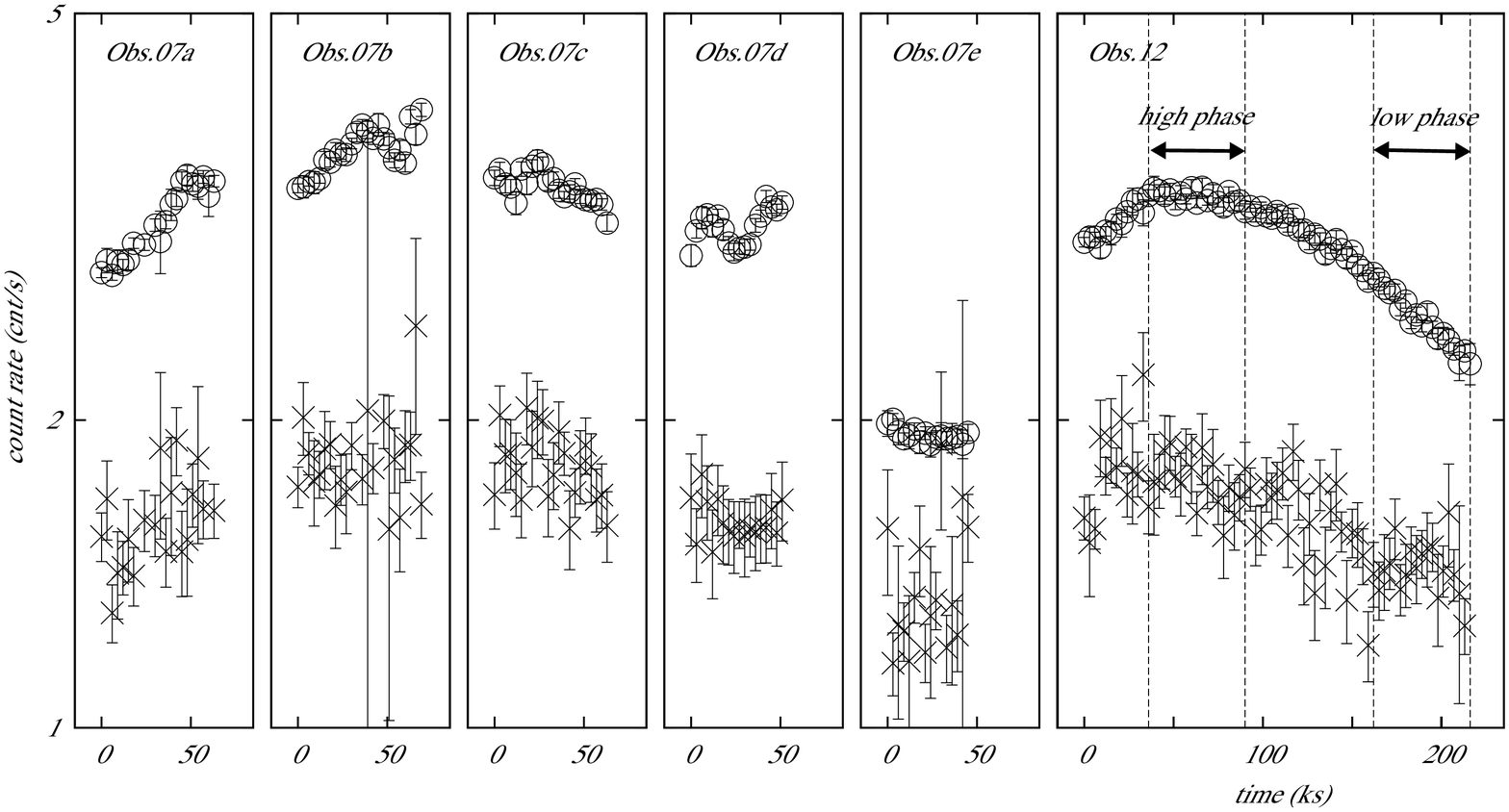}
 \end{center}
 \caption{Background-subtracted and dead-time-corrected light curves of IC4329A, acquired with XIS0 + XIS3 in 2--10~keV (circles) and HXD-PIN (crosses) in 15--45~keV. The latter is multiplied by a factor of 5. In the 2012 light curves, the high and low phases are indicated by arrows.}\label{fig:lc}
\end{figure}

Figure \ref{fig:lc} shows the background-subtracted and dead-time-corrected XIS and HXD-PIN light curves of the individual observations. Thus, the 2--10~keV count rate varied by a factor $\sim 2$ among the 5 observations in 2007, and changed by a factor of $\sim 1.5$ during the 216~ks of gross pointing in 2012. In contrast, the 15--45~keV count rate was less variable.\par
\begin{figure}[h!]
  \begin{minipage}{0.49\hsize}
    \begin{center}
      \includegraphics[width=9.5cm, angle=0]{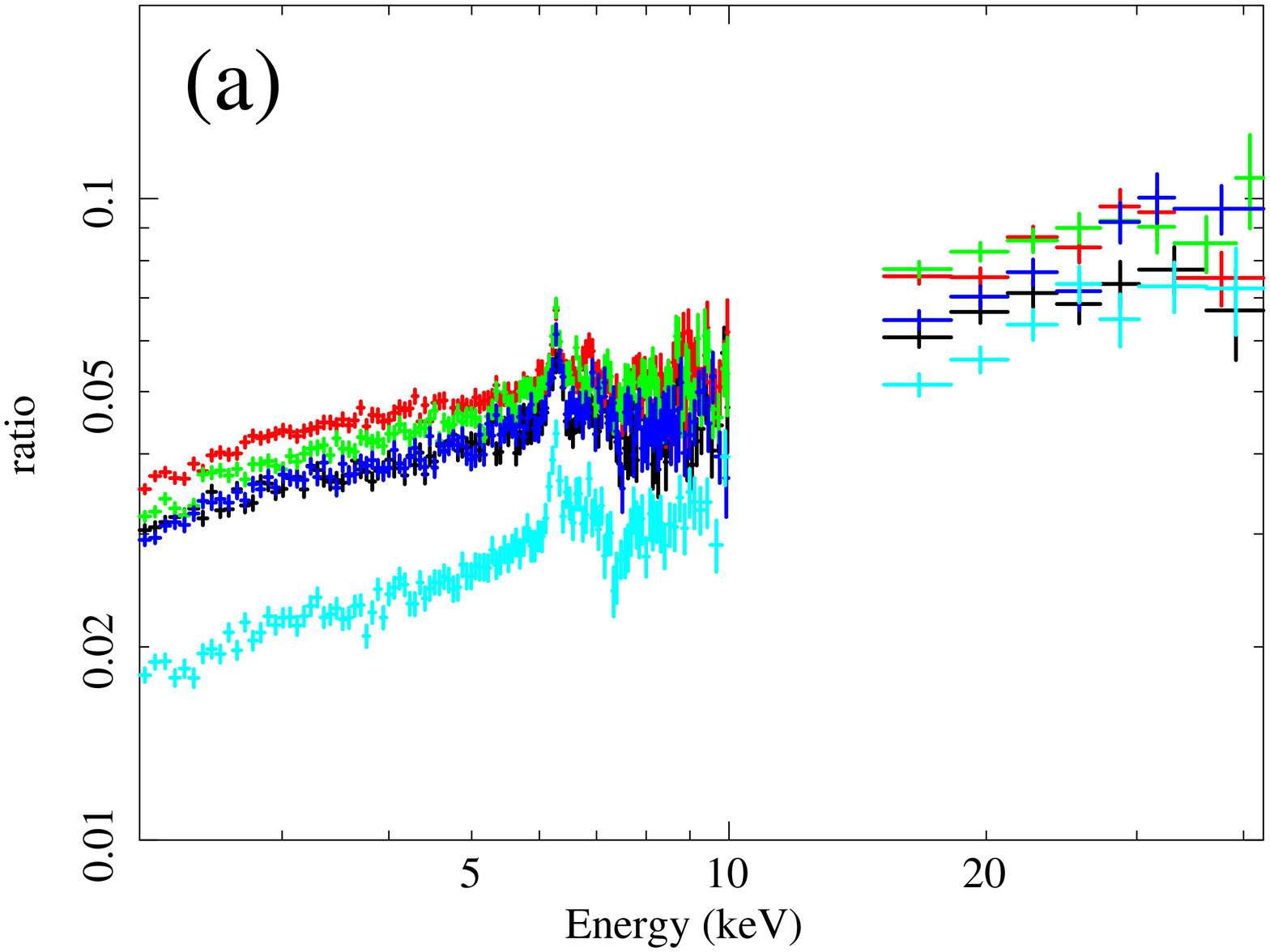}
    \end{center}
  \end{minipage}
  \begin{minipage}{0.49\hsize}
    \vspace{-2mm}
    \begin{center}
      \includegraphics[width=9cm, angle=0]{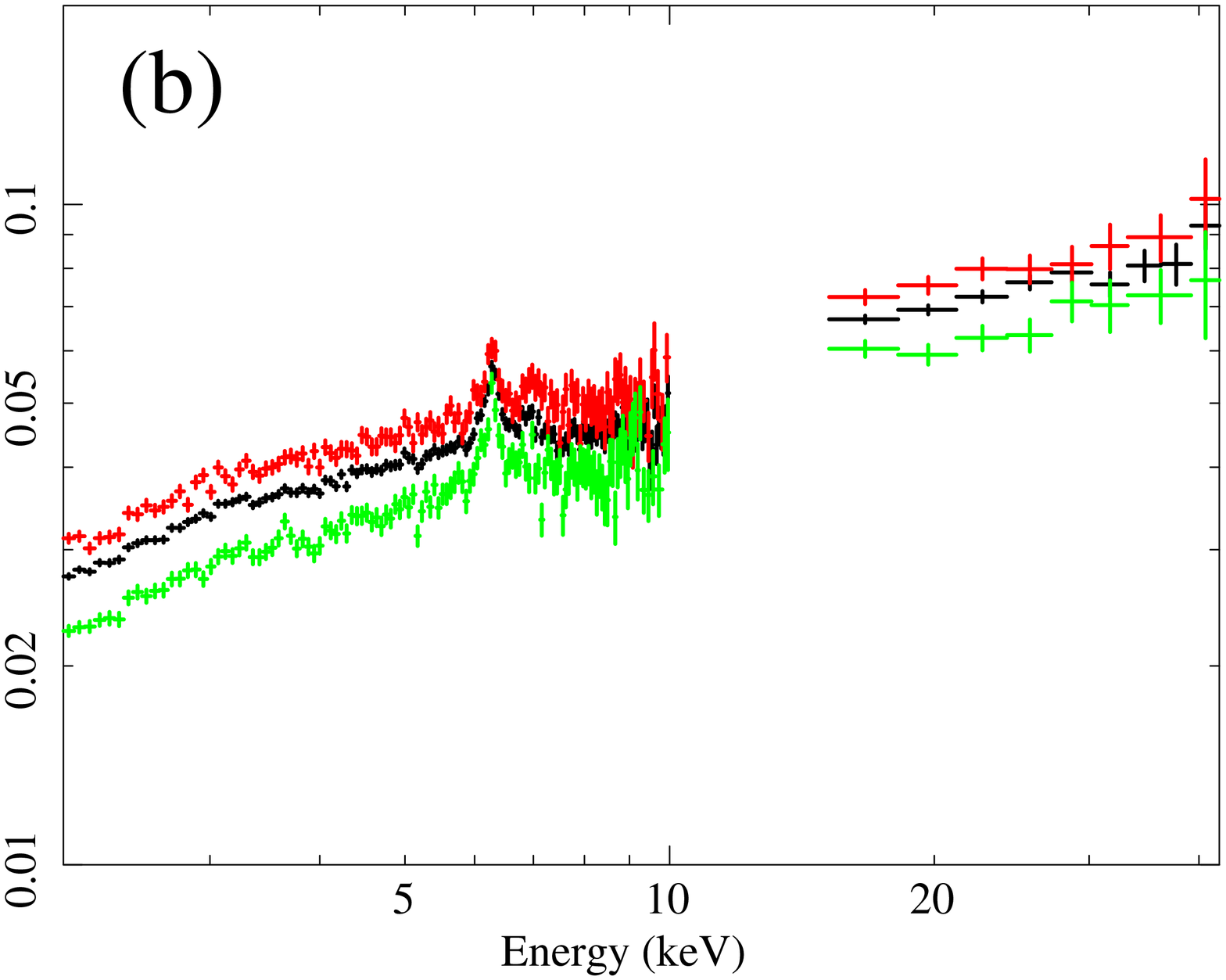}
    \end{center}
  \end{minipage}
  \caption{Background-subtracted and time-averaged spectra of IC4329A obtained with the XIS and HXD-PIN, shown in the form of their ratios to a PL with $\Gamma=2$ and a normalization of 1.0. (a) The spectra obtained in Obs.07a (black), Obs.07b (red), Obs.07c (green), Obs.07d (blue), and Obs.07e (cyan). (b) The spectra of Obs.12 accumulated over the whole observation (black), the high phase (red), and the low phase (green).}\label{fig:spec}
\end{figure}

Background-subtracted spectra of all observations are shown in figure \ref{fig:spec}, in the form of their ratios to a PL model of $\Gamma=2$. This process is similar to converting the spectrum into $\nu F_\nu$ forms, but is less model dependent, because it does not involve deconvolution that requires the data to be fitted with a successful model. Thus, the broad-band continuum exhibited an approximately PL shape of $\Gamma\sim1.7$, with a hint of spectral hardening towards lower intensity. In each spectrum, we can see an Fe K$\alpha$ line and an edge at 6.4~keV and 7.1~keV in the rest frame, respectively. While the PL continuum varied significantly, the Fe K$\alpha$ line was much less variable.

\subsection{Difference spectra}\label{sec:diff}
By taking the difference between a pair of spectra, one when the source is brighter and the other fainter, a component that contributes to the variation of the source will be extracted. To conduct this difference spectrum analysis, we selected Obs.07e (cyan in figure \ref{fig:spec}a) data, when the source was by far the faintest, as the data to be subtracted from the others in 2007. 
Furthermore, as indicated in figure \ref{fig:lc}, we defined 2 segments, namely, the high phase and the low phase, from Obs.12. These segments have nearly the same exposure, and were selected to minimize relative errors of the difference spectrum between them. As presented in figure \ref{fig:spec}b, the high-phase spectrum (red) is 1.5--2 times higher than that in the low-phase. \par
\begin{figure}[h!]
  \begin{minipage}{0.49\hsize}
    \begin{center}
      \includegraphics[width=10cm, angle=0]{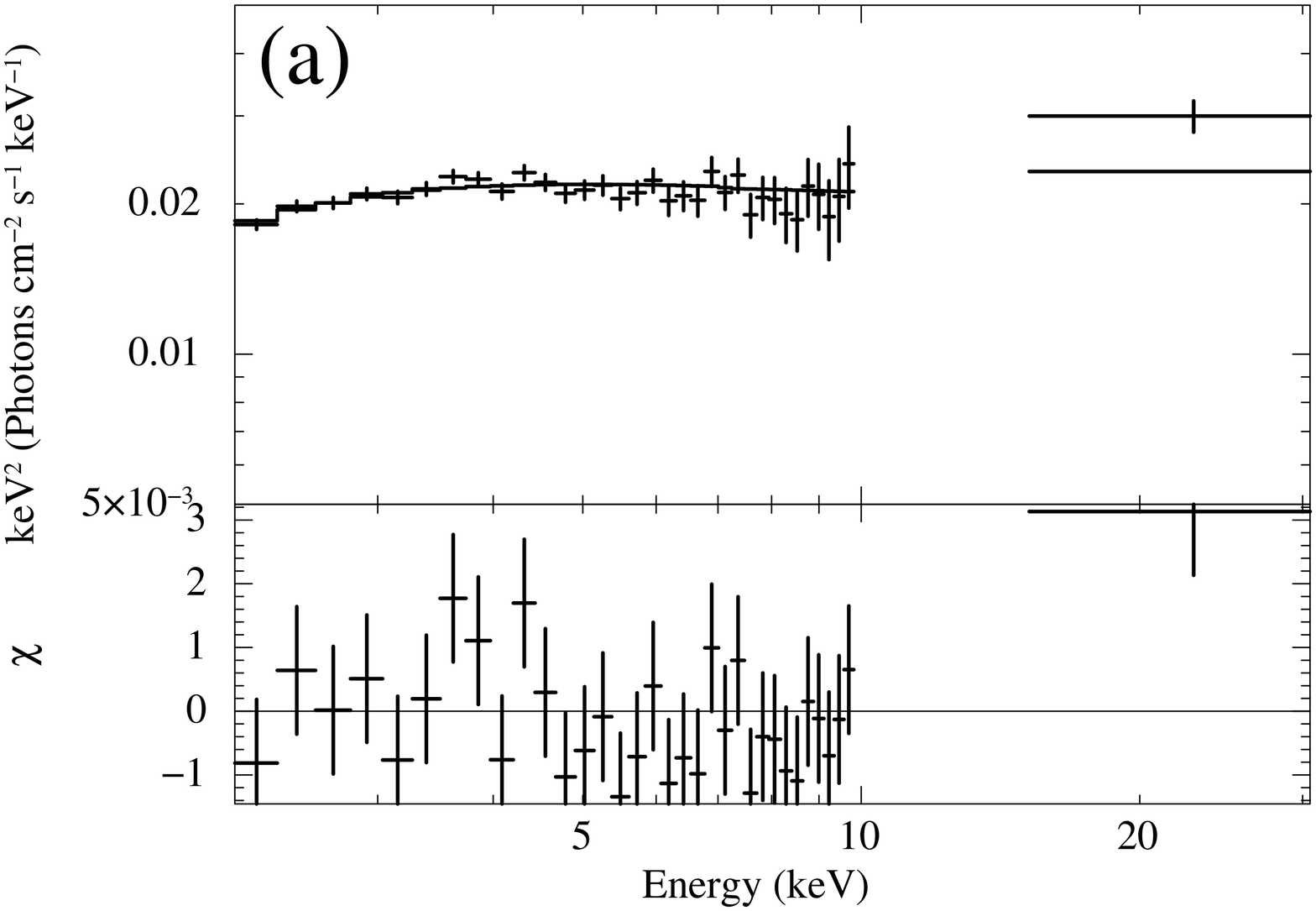}
    \end{center}
  \end{minipage}
  \begin{minipage}{0.49\hsize}
    \vspace{-2mm}
    \begin{center}
      \includegraphics[width=10cm, angle=0]{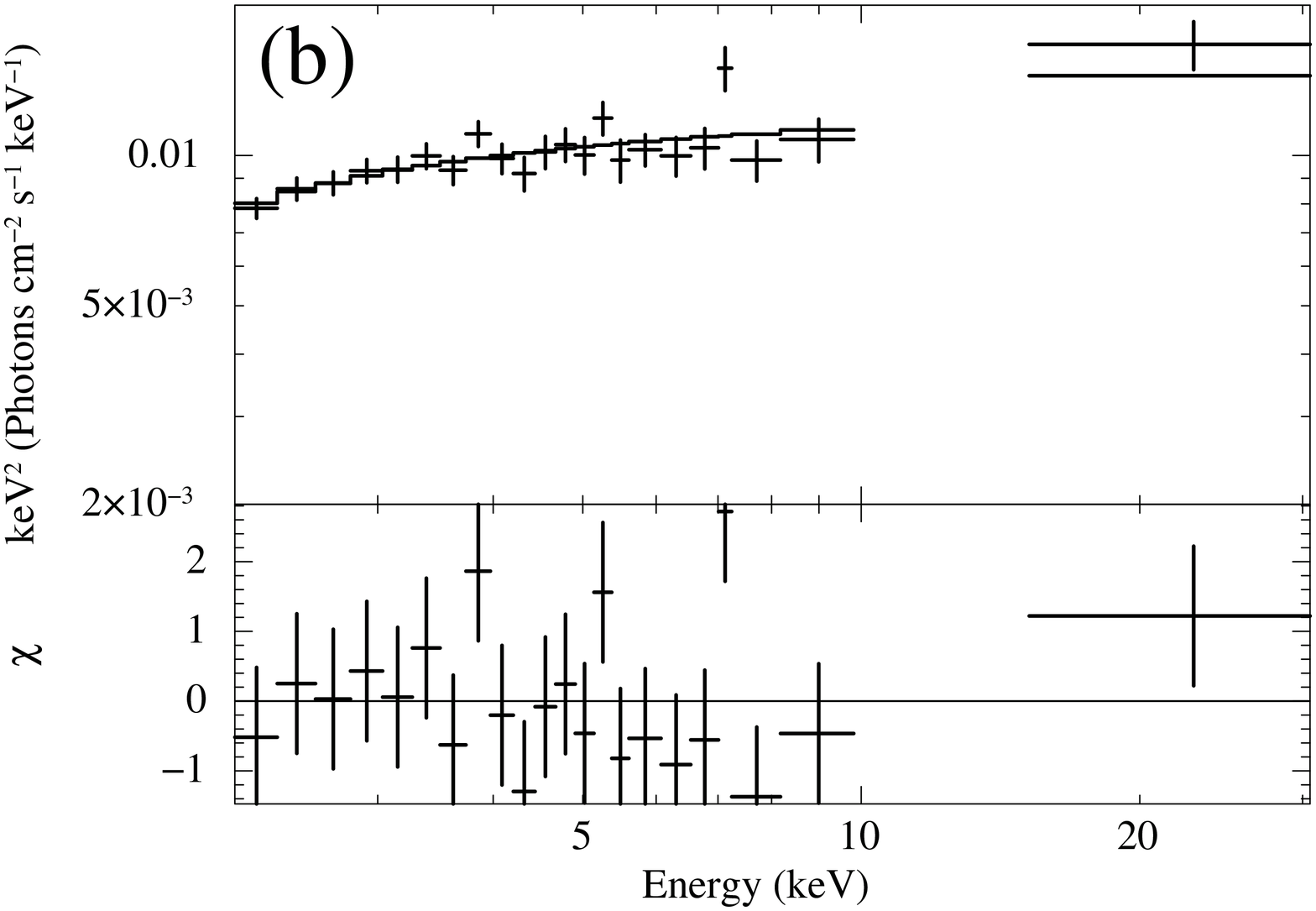}
    \end{center}
  \end{minipage}
 \caption{Difference spectrum between Obs.07b and Obs.07e (panel a), and between the high and low phases in 2012 (panel b), each fitted with an absorbed PL.}\label{fig:diff}
\end{figure}

Figure \ref{fig:diff} shows two examples of difference spectra; one is among the observations in 2007 where Obs.07b (red in figure \ref{fig:spec}a; the brightest) is the ``high" data, and the other within Obs.12. Including these two examples, we produced altogether 5 difference spectra (4 from 2007 and one from 2012), and fitted them with an absorbed PL model, {\tt wabs*powerlaw}. As shown in table \ref{tab:diff}, the fit was all successful, and the derived photon indices were consistent with $\Gamma=2.0$ in all cases. Therefore, the source variation was expressed by intensity changes of a PL-shaped component with a constant slope of $\Gamma\sim2.0$. More specifically, the photon index of the variable PL component was $2.07 \pm 0.04$ (averaged over the four) in 2007, and $1.89^{+0.12}_{-0.11}$ in 2012.\par
Of course, the variability can be more complex, as there still remains a possibility that a single spectral component is varying both in its intensity and shape, or multiple components are varying to jointly produce the observed spectral variability. In these cases, the variations must involve at least two parameters; e.g., the normalization and the slope of the varying single component, or the normalization of the involved two components. However, in order for all the 5 different spectra derived here to have the mutually consistent single shape described by $\Gamma\sim2.0$, a fine-tuning mechanism must be operating among the multiple parameters. Therefore, the presence of a single variable component with insignificant shape changes provides a much more assumption-free and natural interpretation of the difference spectra. A further discussion on this point continues in subsection \ref{sec:c3po}. \par
Since the spectral shape of the variable part of the emission has been quantified, our next step is to examine whether the time-averaged spectra can be reproduced by adding neutral reflection to this $\Gamma\sim2.0$ PL. For this purpose, we fitted the time-averaged spectra of all observations with a simple model, which consists of a PL with the constant photon index $\Gamma=$2.07 (2007) or 1.89 (2012), and its reflection with the abundance fixed at 1 solar and the inclination angle to 60$^{\circ}$; the model form is hence {\tt wabs[0]*(powerlaw[0]+pexmon[0])} in the XSPEC terminology, where ``0" specifies the component number and {\tt pexmon} is a neutral Compton reflection model accompanied by the Fe and Ni lines. The interstellar absorption between the object and the earth is taken into account as {\tt wabs[0]} which is allowed to vary. However, it failed to reproduce the time-averaged spectra with $\chi ^2 / {\rm d.\, o.\, f.\, }\ge 2.3$ (for all the five spectra) in 2007 and $\chi ^2 / {\rm d.\, o.\, f.\, }=3.1$ in 2012. \par
\begin{figure}[h!]
  \begin{center}
    \includegraphics[width=10cm, angle=0]{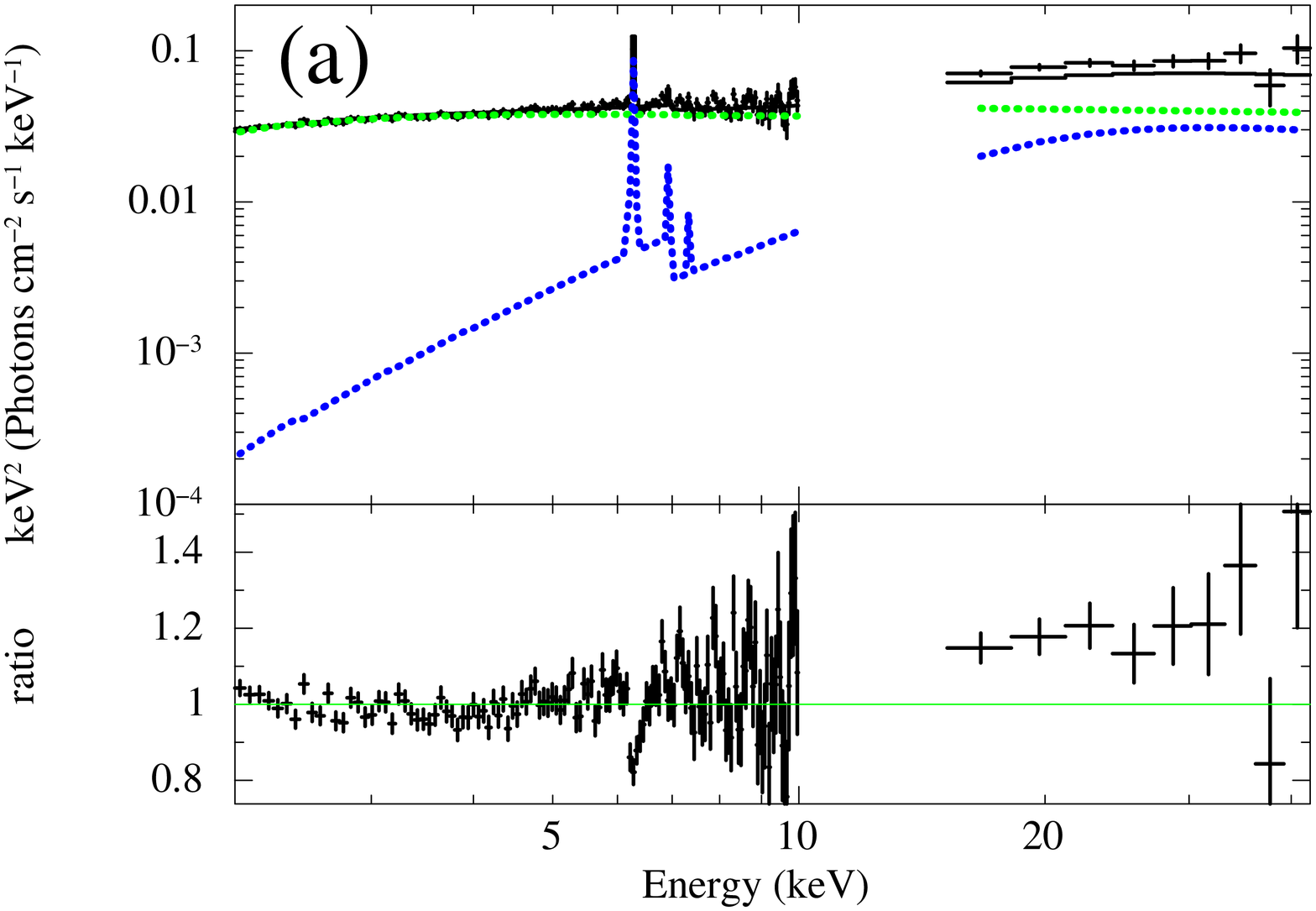}
  \end{center}
  \begin{minipage}{0.49\hsize}
    \begin{center}
      \includegraphics[width=9cm, angle=0]{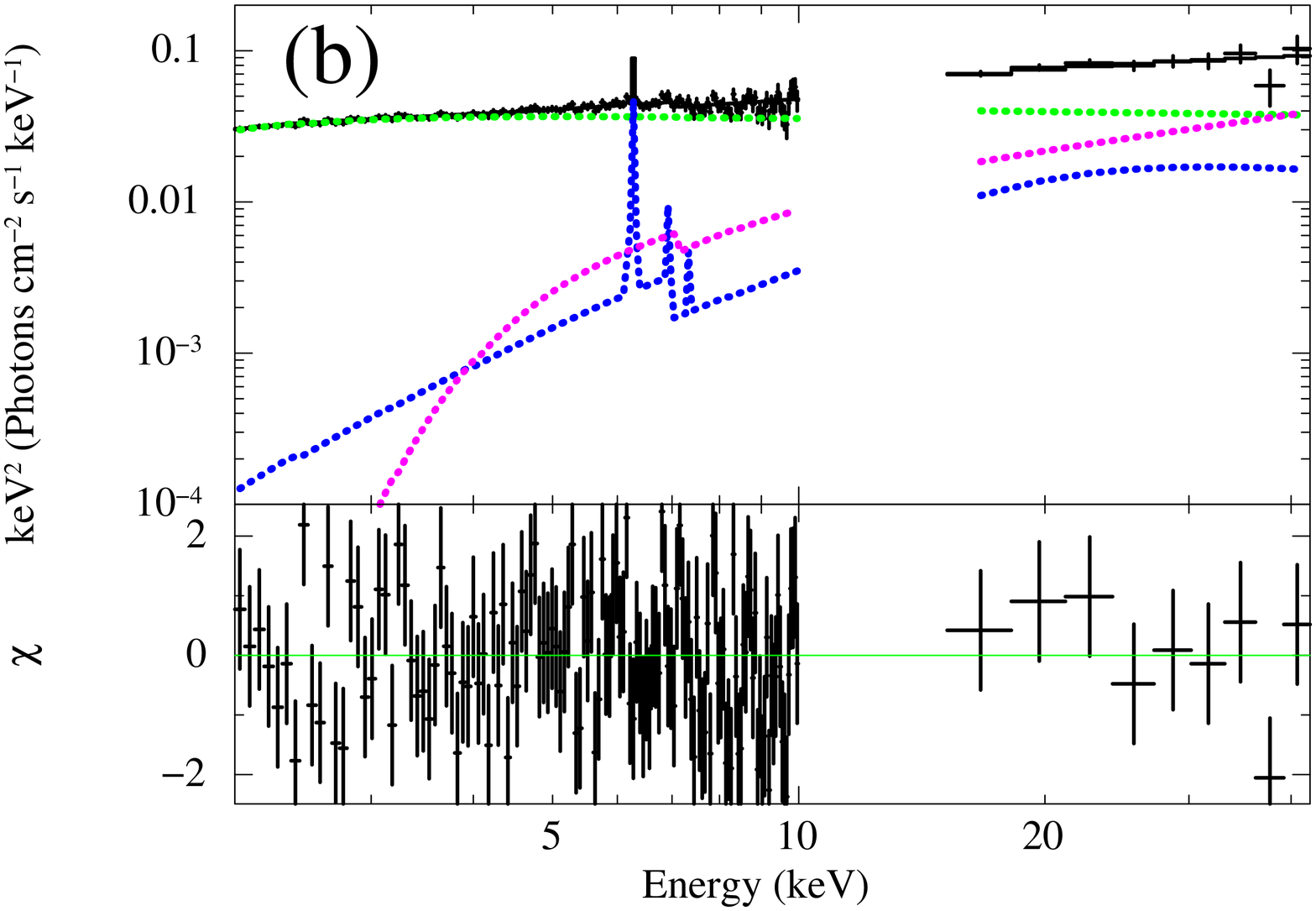}
    \end{center}
  \end{minipage}
  \begin{minipage}{0.49\hsize}
    \vspace{-2mm}
    \begin{center}
      \includegraphics[width=8.2cm, angle=0]{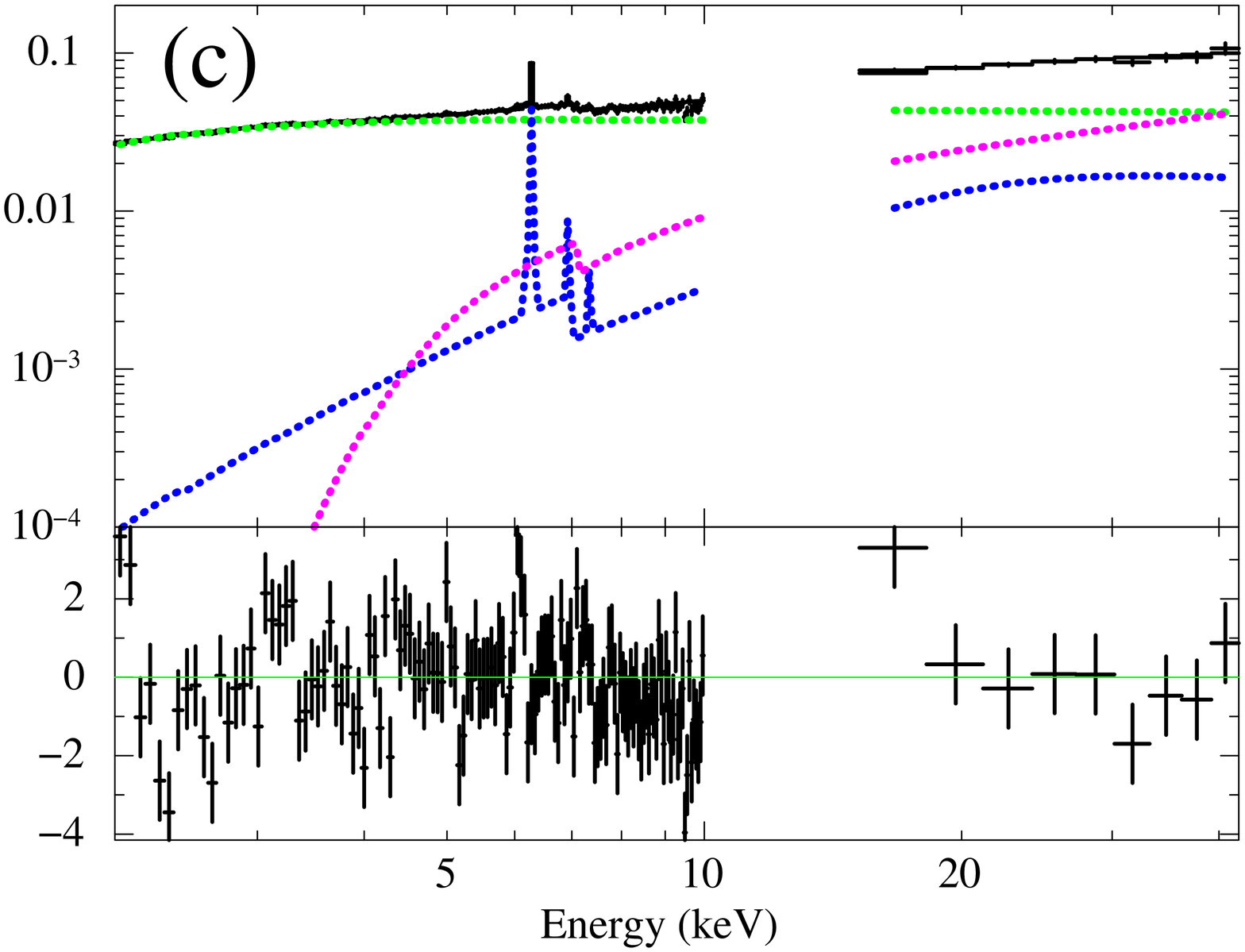}
    \end{center}
  \end{minipage}
 \caption{(a) The time-averaged spectrum of Obs.07a (black), fitted with a model consisting of a fixed $\Gamma=2.07$ PL (green), and its reflection (blue). (b) The same as panel (a) but adding another harder PL (magenta). (c) The same as panel (b), but for Obs.12. In this case, the green PL has a fixed slope of $\Gamma=1.89$.}\label{fig:difffit}
\end{figure}

Taking the case of Obs.07a as an example, let us look into the issue. As shown in figure \ref{fig:difffit}a, the above model failed to reproduce the sharp edge structure at $\sim7.0$~keV, 
and the fit also left significant residuals in the hard X-ray continuum above 15~keV. 
If we try to explain the edge at $\sim7.0$~keV by applying a stronger absorption to the $\Gamma=2$ PL, a column density of $N_{\rm H}\sim8\times10^{22}$~cm$^{-2}$ would be needed. This would strongly attenuate the low-energy continuum, and make the fit completely unacceptable. The problem is not solved even if we allow $\Gamma$ to vary, to incorporate a steeper PL. 
If the inclination of the reflector was allowed to vary, the fit $\chi ^2 / {\rm d.\, o.\, f.\, }$ improved from 353.9/142 to 291.3/141, but the obtained inclination was extreme, $84\fdg2$, and the required solid angle of the reflector was unphysical, $>4\pi$. The fit became nearly acceptable ($\chi ^2 / {\rm d.\, o.\, f.\, }=176.9/139$) when the abundance of the reflector was allowed to vary in addition; the Fe abundance became $0.32\pm0.05$ solar, and that of the other metals was $0.39^{+0.12}_{-0.08}$ solar. This is because the model tried to fill in the deficit at $>15$~keV by increasing the reflection component, but the Fe abundance should be decreased in order not to over-produce the Fe-K line. Obviously, such a low abundance would be unrealistic at the nucleus of a spiral galaxy. In summary, the time-averaged spectrum cannot be reproduced simply by adding the reflection component to the $\Gamma=2.07$ PL. \par
\begin{figure}[h!]
    \begin{center}
      \includegraphics[width=8cm, angle=0]{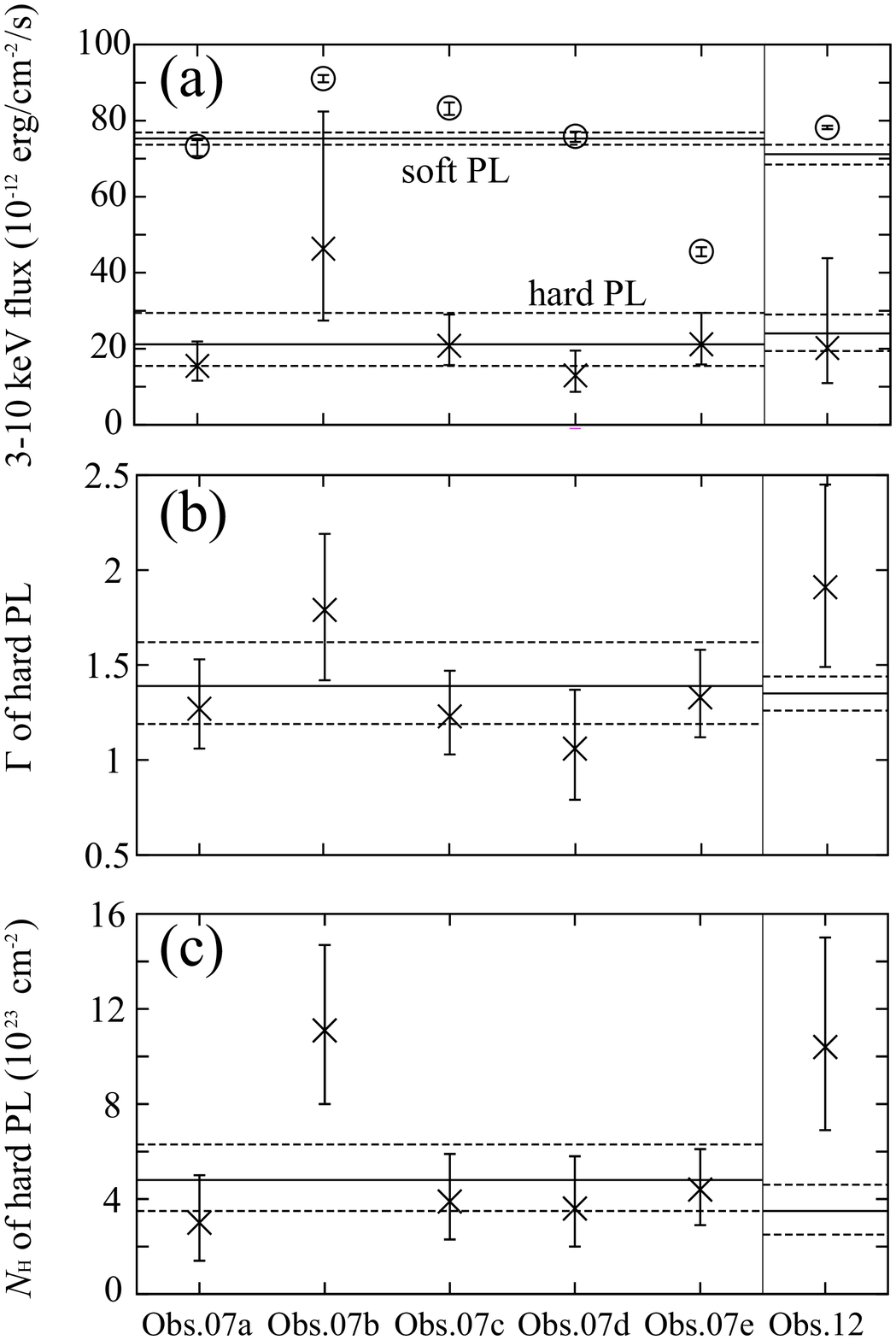}
    \end{center}
 \caption{The best-fit parameters obtained with the ``two PLs plus one reflection" model. Data points with error bars specify those obtained with the difference spectrum method. Solid lines represent the mean values derived with the C3PO method, with the upper/lower limits indicated by dashed lines. (a) The 3--10~keV flux of the soft (circle) and hard (cross) PL before absorption. (b) $\Gamma$ of the hard PL. (c) $N_{\rm H}$ of the hard PL.}\label{fig:par}
\end{figure}

From the above results, the time-averaged spectra require at least one more component to be added to the model. The component must have a rather hard slope to explain the 15--45~keV excess seen in figure \ref{fig:difffit}a, and an Fe-K edge to reproduce the negative structure at $\sim7.0$~keV, but no strong Fe-K emission line. The simplest of such a component would be an absorbed hard PL, namely, {\tt wabs[1]*powerlaw[1]}. As shown in figure \ref{fig:difffit}b and figure \ref{fig:difffit}c, the addition of this new spectral component to the above simplest model has significantly improved the fit to the time-averaged spectrum, to $\chi ^2 / {\rm d.\, o.\, f.\, }\le1.3$ in all the 2007 observations and to $\chi ^2 / {\rm d.\, o.\, f.\, }= 1.84$ in 2012. Figure \ref{fig:par} shows the best-fit parameters obtained by this ``two PLs plus one reflection" model, which consists of the $\Gamma=2.07$ PL, the newly added harder PL, and the reflection of the former PL with the solar abundance and the inclination angle of 60$^{\circ}$. As seen from figure \ref{fig:par}a, the source variation in 2007 (particularly the marked decrease in Obs.07e) is mostly carried by the soft PL, with the hard PL contributing $\lesssim33$\% of the overall variability. This is also the case in 2012, because the difference spectrum (figure \ref{fig:diff}b) is soft. (However, in relative sense, the hard PL could vary by up to a factor of $\sim2$, which is close to that of the soft PL.) \par
The hard PL was found to have $\Gamma\sim1.4$ in all observations. As a result, the emission from IC4329A has been successfully explained by incorporating the three components; the highly-variable soft PL component with $\Gamma\sim2.0$, its reflection accompanied by the Fe K$\alpha$ line, and the less-variable harder PL with $\Gamma=1.4\pm0 .1$. The absorption working on the last component has a rather high column density as $N_{\rm H}=(5.0^{+0.9}_{-0.7})\times 10^{23}$~cm$^{-2}$, as determined by the Fe-K edge in the data. Although $N_{\rm H}$ for the hard PL in Obs.07b and Obs.12 are inconsistent with the others within 90\% confidence errors, they become consistent at 98\% confidence level. Even though the high $N_{\rm H}$ strongly attenuates the $\lesssim4$~keV part of the hard PL, the effect is not apparent in the time-averaged spectra, since the hard-PL flux therein is $\lesssim0.3$ times that of the softer PL even if the absorption were neglected. \par

\begin{table}
  \caption{Best-fit parameters of the difference spectra by an absorbed PL model.}\label{tab:diff}
  \begin{center}
    \begin{tabular}{llllll}
      \hline
      $({\rm high}-{\rm low})$ & ratio$^*$ & $N_{\rm H}^{\dagger}$ & $\Gamma$ & $\chi ^2 / {\rm d.\, o.\, f.\, }$\\ \hline
      $({\rm Obs.07a}-{\rm Obs.07e})$ & 1.59 & $0.84^{+0.40}_{-0.38}$ & $2.13\pm0.10$ & 31.96/34\\
      $({\rm Obs.07b}-{\rm Obs.07e})$ & 1.87 & $0.81\pm0.27$ & $2.08\pm0.07$ & 33.15/34\\
      $({\rm Obs.07d}-{\rm Obs.07e})$ & 1.75 & $0.77\pm0.29$ & $1.93^{+0.08}_{-0.07}$ & 24.93/34\\
      $({\rm Obs.07e}-{\rm Obs.07e})$ & 1.60 & $1.09^{+0.40}_{-0.39}$ & $2.14^{+0.10}_{-0.11}$ & 32.56/34\\
      Obs.12 $({\rm high}-{\rm low})$ & 1.30 & $0.58^{+0.49}_{-0.47}$ & $1.89^{+0.12}_{-0.11}$ & 33.0/31\\
      \hline
    \end{tabular}
  \end{center}
  \begin{itemize}
    \item[$^*$] Ratio of fluxes in 2--10~keV. 
    \item[$^\dagger$] The absorbing column density in (10$^{22}$~cm$^{-2}$). 
  \end{itemize}
\end{table}

\subsection{Count-Count Correlation with Positive Offset method}\label{sec:c3po}
Now that we have successfully accomplished the difference spectrum analysis in subsection \ref{sec:diff}, let us examine whether the C3PO method (section \ref{sec:intro}; Noda et al. \yearcite{Noda2011b}, \yearcite{Noda2013b}) gives the same results. This cross check is necessary, because the two methods are not completely equivalent. For example, the difference spectrum technique gives an answer as long as the source is variable, but there may still be a possibility that the source variations are caused by some mechanisms other than the intensity changes of a single component of a fixed shape. 
This intrinsic risk can be avoided by using the C3PO method, because we can test that assumption by examining whether the underlying CCPs have a good linearity or not. Furthermore, our choice of the absorbed PL in subsection \ref{sec:diff}, to bring the fit acceptable, was somewhat arbitrary; this is another point where the C3PO method can give a more assumption-free answer. Yet another merit of the C3PO method is that the intensities of the varying spectrum and the non-varying part can be determined in their absolute sense, unlike in the difference spectrum analysis. \par
Given the above consideration, is there conversely any merit of using the difference spectrum method? In fact, the C3PO technique also has one fundamental weak point: the assumption that the stationary flux is negligible in the reference band cannot be justified within its own framework. 
However, this difficulty can be solved if the difference spectrum method is employed, because it can determine the shape of the variable component without any assumption. 
Actually, the fits in figure \ref{fig:difffit}a and figure \ref{fig:difffit}b, incorporating the difference spectra, imply that the stationary signal is $\lesssim1$\% of the overall signal in the 2--3~keV band. This is because the 2--3~keV spectral shape of the time-averaged spectrum is almost identical to that of the difference spectrum (i.e., the variable PL with $\Gamma\sim2.0$). Thus, no other components are needed in this range. Hence, we are justified to use the 2--3~keV range as our reference band. Furthermore, as already shown by \citet{Noda2013a}, the C3PO results are not changed in essence, especially in deriving the variable component, even if the reference band contains some amount of non-varying signals. \par
\begin{figure}[h!]
  \begin{center}
    \includegraphics[width=16cm]{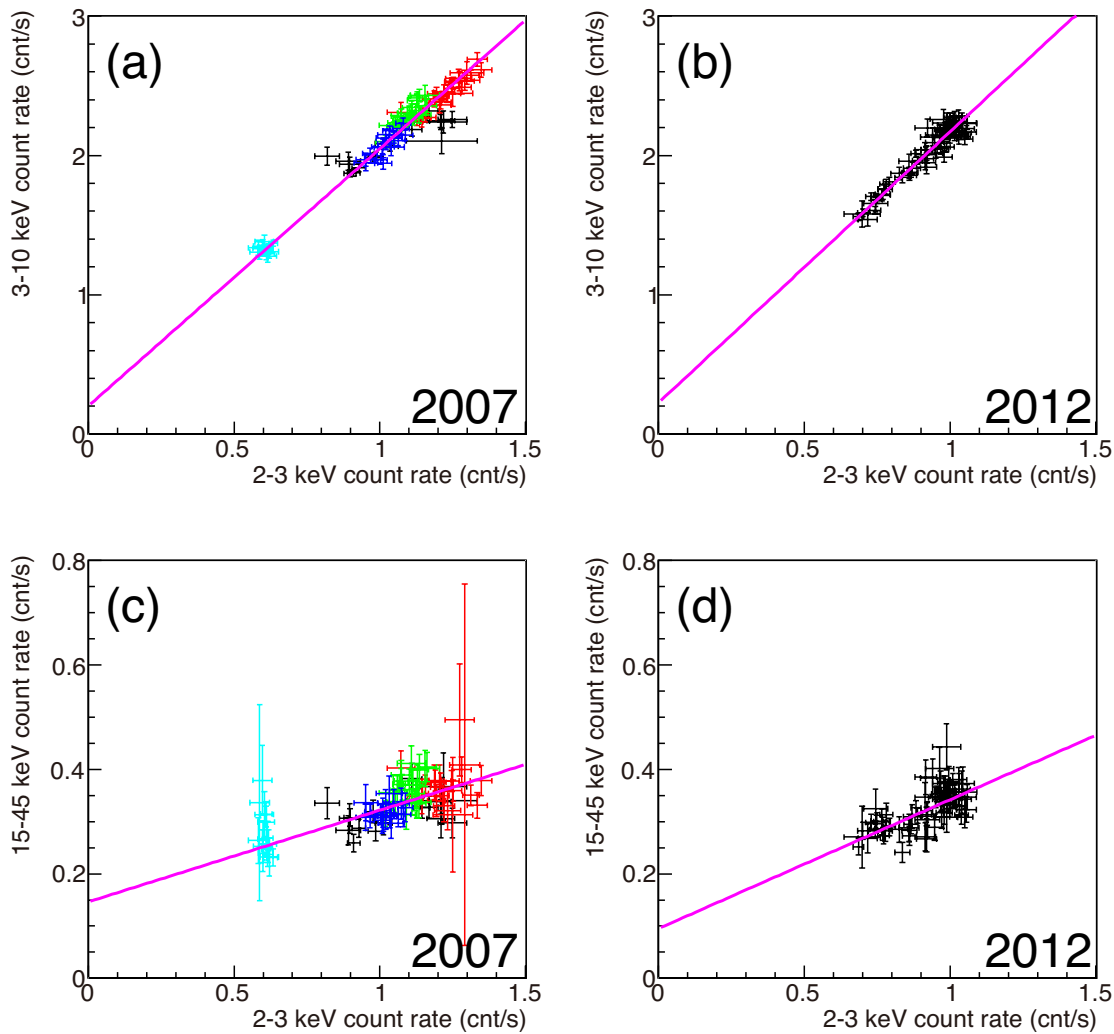}
  \end{center}
 \caption{
CCPs of the present Suzaku data. Panels (a) and (b) are those between the reference 2--3~keV and the broad XIS band (3--10~keV), whereas (c) and (d) are those between the 2--3~keV and the broad HXD-PIN band (15--45~keV). Panels (a) and (c) are from 2007, where the colors specify the observations as in figure \ref{fig:spec}a, while (b) and (d) are from 2012. In all the plots, the best-fit line of equation (\ref{eq:c3po}) is shown in magenta. }\label{fig:ccp}
\end{figure}

Figure \ref{fig:ccp} shows 2--3~keV vs. 3--10~keV and 2--3~keV vs. 15--45~keV CCPs, incorporating all the present observations of IC4329A (but separately for the 2007 and 2012 data). Thus, in all the CCPs, the data points exhibit approximately one-dimensional distributions, implying that the source variability can be described by one parameter. This agrees with our argument in subsection \ref{sec:diff}. Furthermore, in the XIS band with high statistics, the data distributions are very linear, except a few data points in Obs.07a. This supports the assumption considered in subsection \ref{sec:diff}, that the source varies when a single component with a constant shape changes its intensity. Then, according to the recipe of the C3PO method, the data points in each CCP were fitted with a straight line expressed by 
\begin{equation}
  y=Ax+B,
  \label{eq:c3po}
\end{equation}
in which $x$ is the count rate in the 2--3~keV reference band, and $y$ is that in the relevant energy band. Thus, $y$ can be decomposed into a variable part, $A\overline{x}$, and a stationary part, $B$, where $\overline{x}$ is the average count rate in the reference band. In Appendix, we consider an alternative a case wherein the variation is caused by {\it slope} changes of a single component. \par
\begin{figure}[h!]
  \begin{center}
    \includegraphics[width=10cm, angle=0]{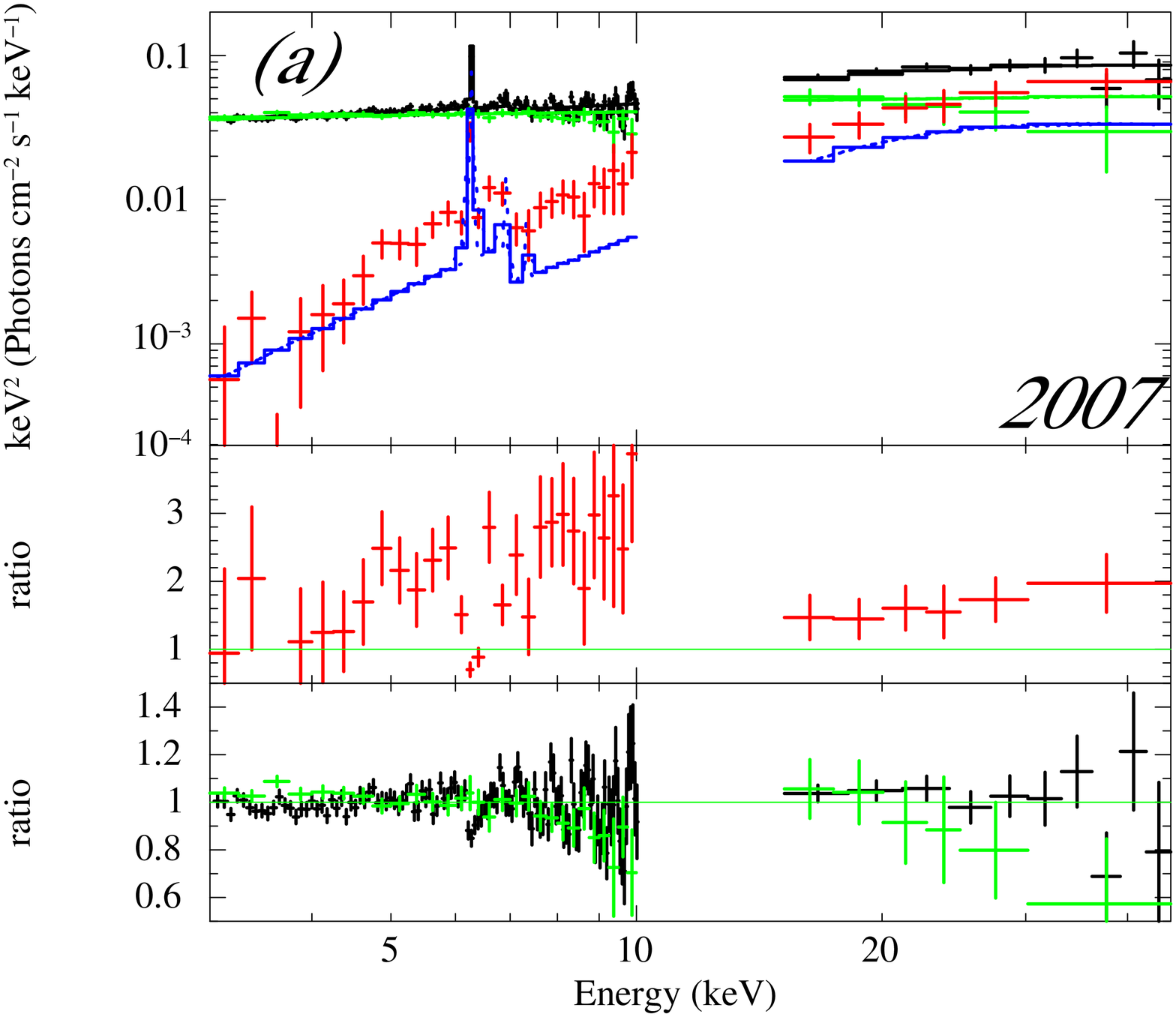}
  \end{center}
  \begin{minipage}{0.49\hsize}
    \begin{center}
      \includegraphics[width=9cm, angle=0]{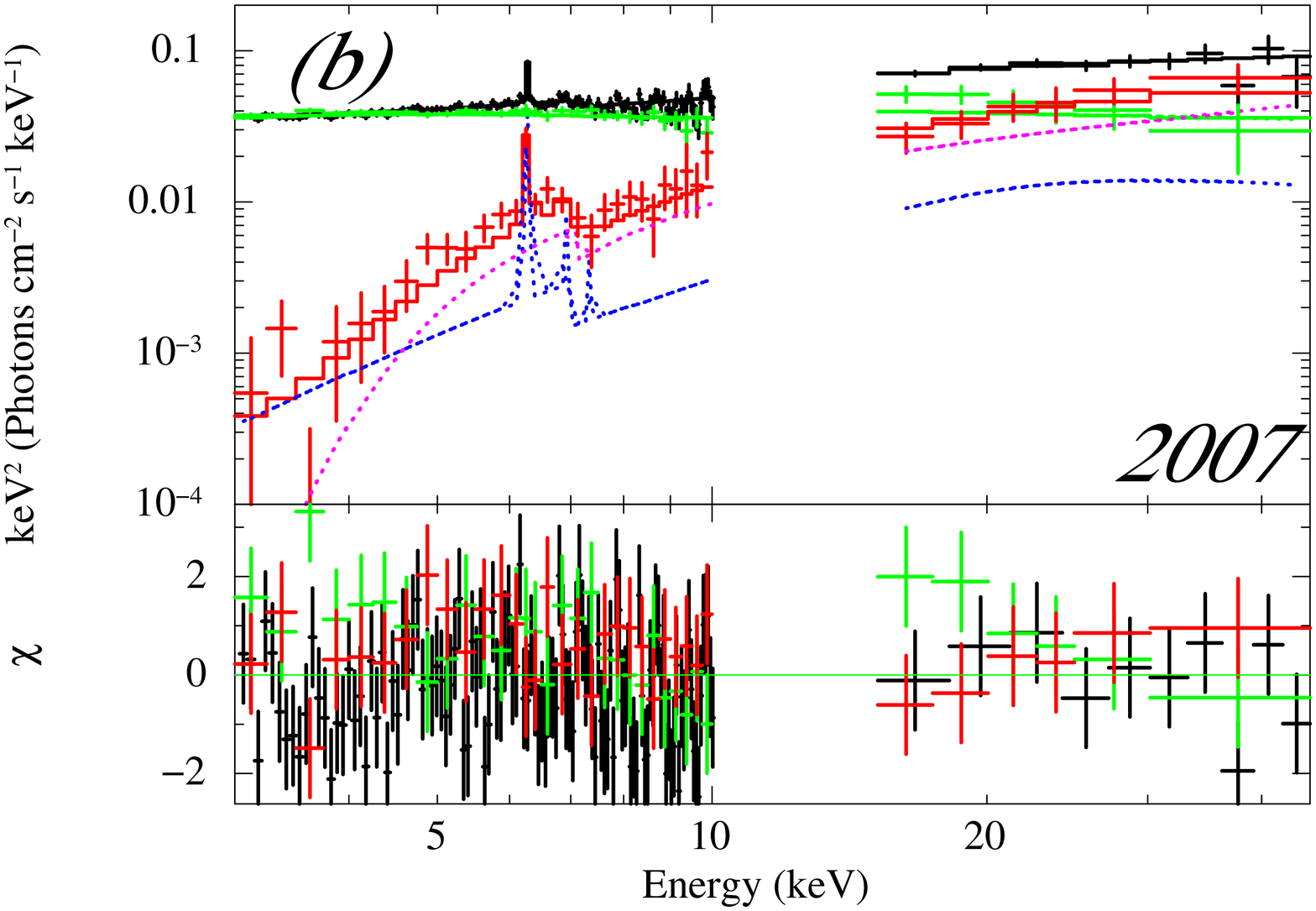}
    \end{center}
  \end{minipage}
  \begin{minipage}{0.49\hsize}
    \vspace{-2mm}
    \begin{center}
      \includegraphics[width=8.2cm, angle=0]{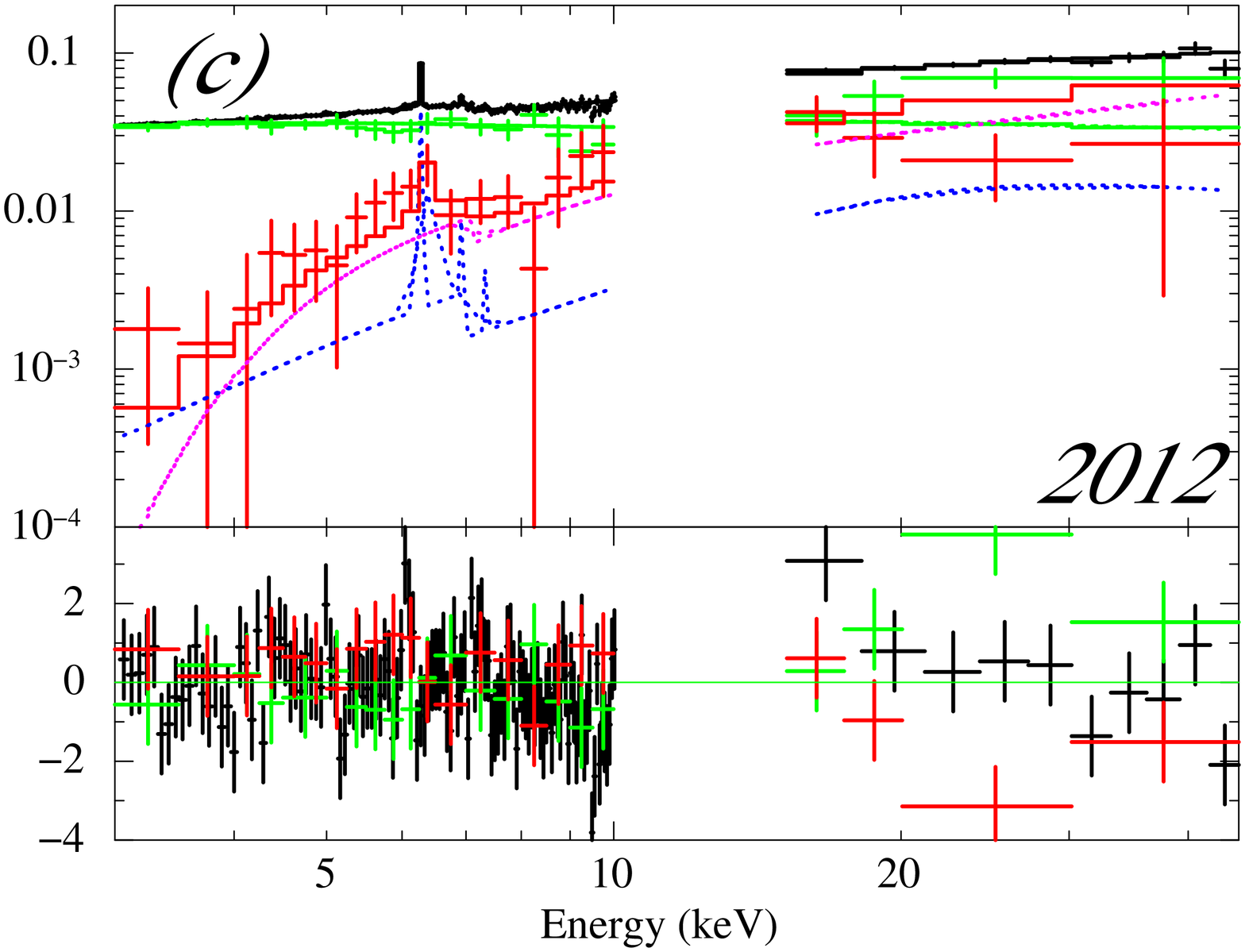}
    \end{center}
  \end{minipage}
  \begin{minipage}{0.49\hsize}
    \begin{center}
      \includegraphics[width=9cm, angle=0]{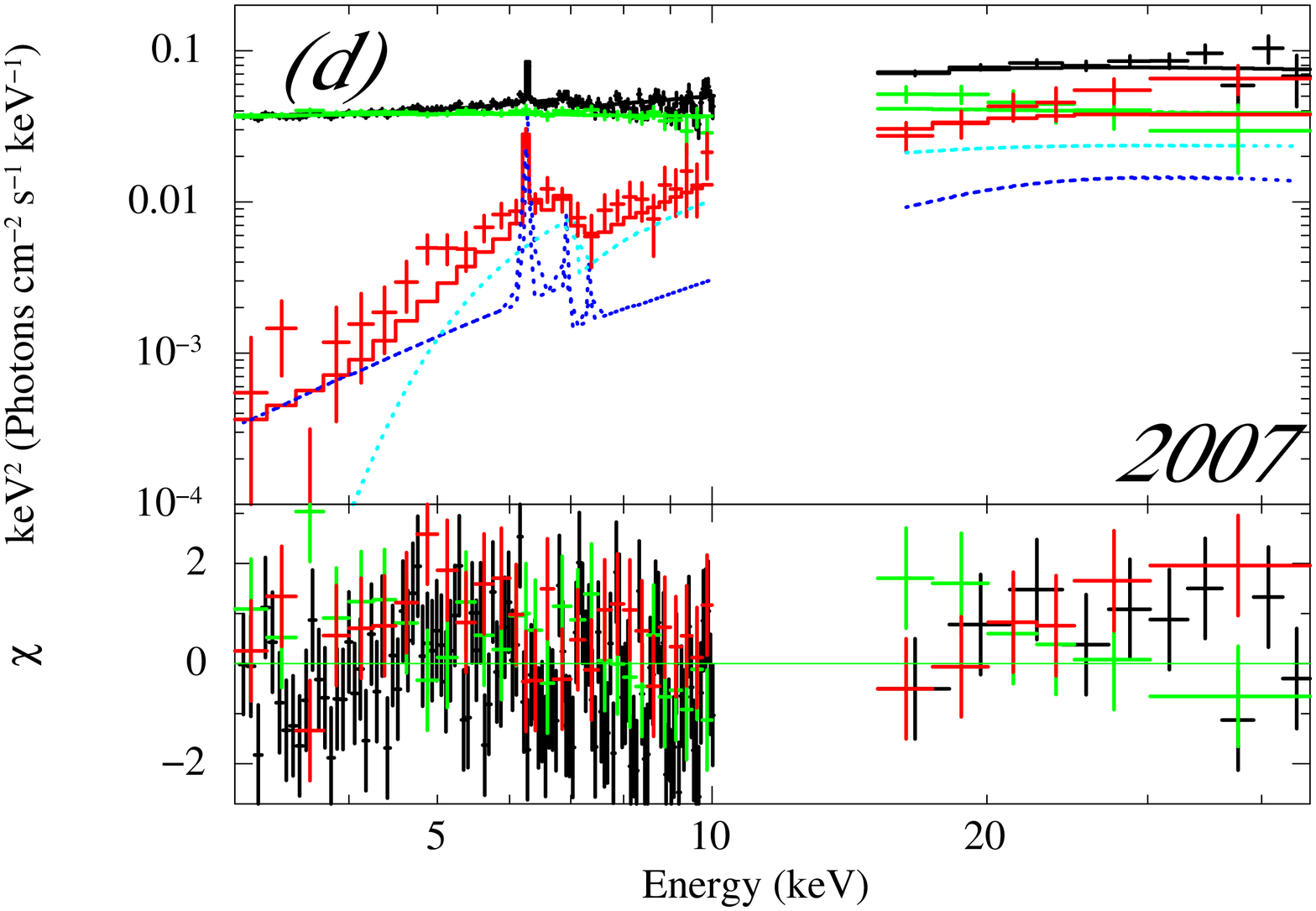}
    \end{center}
  \end{minipage}
  \begin{minipage}{0.49\hsize}
    \vspace{-2mm}
    \begin{center}
      \includegraphics[width=8.2cm, angle=0]{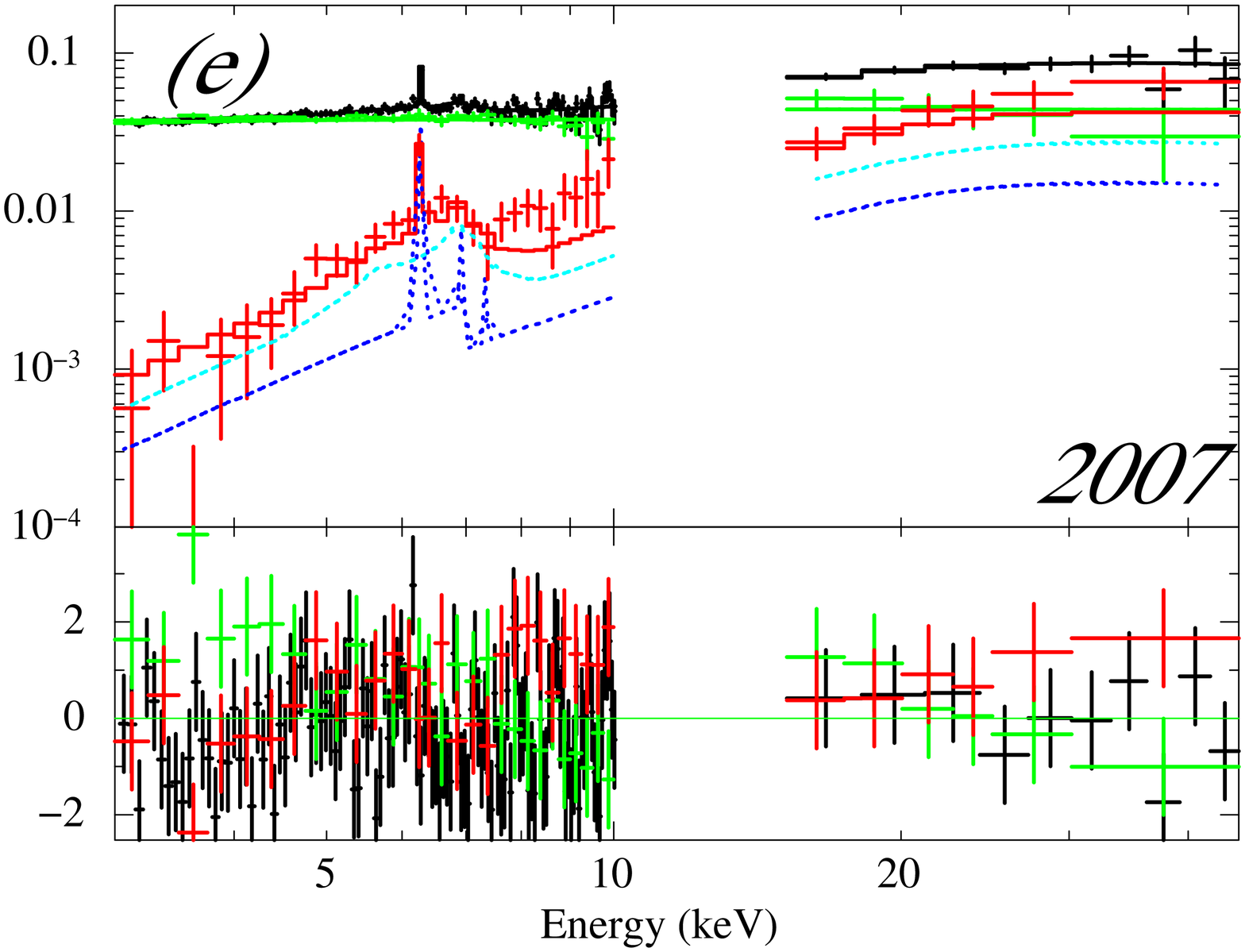}
    \end{center}
  \end{minipage}
%  \begin{center}
%    \includegraphics[width=10cm, angle=0]{20151008_c3po_2refl.eps}
%  \end{center}
 \caption{(a) The spectrum averaged over all the 2007 observations (black), shown together with the variable part (green) and the stationary part (red) extracted with the C3PO method. The variable and stationary parts are fitted with {\tt model\_V} and {\tt model\_S}, respectively, while the averaged spectrum with the sum of {\tt model\_V} and {\tt model\_S}. Here, {\tt model\_V} consists of a single PL (green), and {\tt model\_S} consists of neutral reflection (blue). The residuals are given in ratios, and separately in the bottom tow panels. (b) The same as panel (a) but adding a strongly-absorbed harder PL (magenta) to {\tt model\_S}. (c) The same as panel (b) but in 2012. (d)(e) The same as panel (b) but replacing the hard PL with the partial covering absorption component (panel d; cyan) and the relativistic reflection (panel e; cyan).}\label{fig:c3po}
\end{figure}

Now that the CCP linearity has been confirmed using the broad-band data, we subdivided the 3--10~keV XIS range into 18 finer bands, and the 15--45~keV HXD-PIN coverage into 3 finer bands as well, to create altogether 21 CCPs. In each CCP, the data points have been fitted by equation (\ref{eq:c3po}) with $\chi ^2 / {\rm d.\, o.\, f.\, }<1.4$. 
Since $A\overline{x}$ and $B$ in equation (\ref{eq:c3po}) again specifies the variable and stationary parts of the emission in each energy band, respectively, we can collect the values of $Ax$ and $B$ over the 21 bands to construct a variable spectrum and a stationary spectrum, respectively. The results of this decomposition are presented in figure \ref{fig:c3po}a. In panel (a), the black data points indicate the spectrum averaged over all the 2007 observations, while the green and red points are their variable and stationary parts, respectively, derived via the C3PO decomposition. By definition, the green and red spectra add up to become identical to the black one. In the same way, the 2012 time-averaged spectrum was decomposed into their variable part (green) and stationary part (red), as shown in figure \ref{fig:c3po}c. Thus, the stationary part is characterized by a very hard slope, a strong low-energy decrease, a narrow Fe-K line, and a strong Fe-K edge. \par
In figure \ref{fig:c3po}, the errors of the variable and stationary parts are defined as 68\% confidence fitting errors to the CCPs. As \citet{Noda2013b} explained, these errors are larger than simply splitting the statistical errors of the total emission into those of the two constituent parts, because errors in $A$ (the CCP slope) and $B$ (the CCP intercept) are rather strongly coupled with each other. This mates the simultaneous fit, to be described below, more conservative. \par
After decomposing a time-averaged spectrum, the C3PO method further performs a simultaneous model fit, employing a pair of models denoted as {\tt model\_V} and {\tt model\_S}. The variable part is fitted with {\tt model\_V}, the stationary part with {\tt model\_S}, and the time-averaged spectrum with ${\tt model\_V}+{\tt model\_S}$, all jointly. Since the time-averaged spectra have high statistics, a systematic error of 1\% was added to them. Like in the first step of difference spectrum analysis, we first used {\tt wabs[0]*powerlaw[0]} as {\tt model\_V}, and the reflection component as {\tt model\_S}; the latter is motivated by the shapes of the stationary spectra. As shown in figure \ref{fig:c3po}a, however, the fit was  unacceptable with $\chi ^2 / {\rm d.\, o.\, f.\, }=402.6/196$; the residuals of the stationary part and the time-averaged spectrum are similar to those of figure \ref{fig:difffit}a. Attempts to improve the fit, by allowing the inclination and the abundance of the reflector, gave nearly the same results as in subsection \ref{sec:diff}. These fits failed because the stationary spectrum bears an Fe-K line that can be reasonably attributed to the reflection component, but exhibits too high a hard X-ray continuum to be explained in that way. In short, {\tt model\_S} cannot be expressed by a reflection component alone. \par
Based on the above results, we again added the {\tt wabs[1]*powerlaw[1]} term to {\tt model\_S}, to find that the simultaneous fit became acceptable, as given in figure \ref{fig:c3po}b, with $\chi ^2 / {\rm d.\, o.\, f.\, }=228.4/193$. 
In Obs.12, similarly, the simultaneous fit became much better by adding the second PL to {\tt model\_S}, from $\chi ^2 / {\rm d.\, o.\, f.\, }=374.7/171$ to $\chi ^2 / {\rm d.\, o.\, f.\, }=199.7/168$ (figure \ref{fig:c3po}c). By these fits, $\Gamma$ of the variable soft PL was determined as $\Gamma = 2.12\pm0.06$ in 2007 and $\Gamma = 2.12\pm0.07$ in 2012. The former $\Gamma$ is consistent with that derived in subsection \ref{sec:diff} within 90\% confidence limits, and the latter is within 95\% confidence limits. The spectral parameters of the constituent components derived with the C3PO method are summarized in table \ref{tab:par}. Those of the {\tt wabs[1]*powerlaw[1]} component are shown in figure \ref{fig:par} as horizontal lines, in comparison with those from the difference spectrum method. Including this comparison, the C3PO method has given the results which are consistent with those from the difference spectrum analysis. As a result, the presence of the harder PL with $\Gamma\sim1.4$ has been confirmed with both these two methods. \par

\begin{table}
  \caption{Best-fit parameters of the simultaneous fitting by ``two PLs plus one reflection" model.}\label{tab:par}
  \begin{center}
    \begin{tabular}{lllllllll}
      \hline
      data & $\Gamma$ (soft) & $\Gamma$ (hard) & $N_{\rm H}^*$ & flux$^\dagger$ (soft) & flux$^\dagger$ (hard) & solid angle$^\ddagger$ & $\chi ^2 / {\rm d.\, o.\, f.\, }$\\ \hline
      2007 & $2.12\pm0.06$ & $1.39^{+0.23}_{-0.20}$ & $4.8^{+1.5}_{-1.3}$ & $7.53\pm0.16$ & $2.12^{+0.82}_{-0.57}$ & $(0.80\pm0.14)\pi$ & 228.4/193\\
      2012 & $2.12\pm0.07$ & $1.35\pm0.09$ & $3.5^{+1.1}_{-1.0}$ & $7.12^{+0.25}_{-0.27}$ & $2.40^{+0.50}_{0.46}$ & $(1.32\pm0.14)\pi$ & 199.7/168\\
      \hline
    \end{tabular}
  \end{center}
  \begin{itemize}
    \item[$^*$] The absorbing column density of the hard PL in (10$^{23}$~cm$^{-2}$). 
    \item[$^\dagger$] The 3--10~keV fluxes in $10^{-11}$~erg~cm$^{-2}$~s$^{-1}$.  
    \item[$^\ddagger$] The solid angle of the reflector. 
  \end{itemize}
\end{table}

\section{Discussion}

\subsection{Summary of the results}
We analyzed the Suzaku data of IC4329A acquired in 2007 (on 5 occasions) and 2012, to decompose the X-ray spectra with the difference spectrum analysis as well as the C3PO method. In both these analyses (particularly the C3PO method), the 2--45~keV source variation on time scales of days to weeks were successfully interpreted as intensity changes of the single component with the constant shape. As a result, the 3--45~keV spectra have been decomposed into the variable $\Gamma\sim2.0$ PL, its reflection, and the stationary harder PL with $\Gamma\sim1.4$. 
The two methods not only gave consistent decompositions, but also strengthened the result in a complementary way. Namely, the difference spectrum analysis confirmed that the reference band (2--3~keV in this particular case) has almost no stationary signals, and the C3PO method revealed that signals in all energy bands are linearly correlated with those in the 2--3~keV range. We have thus confirmed the presence of the $\Gamma\sim1.4$ hard component which contributes to the stationary signals. \par

\subsection{Alternative interpretations}
In previous studies of Type I Seyferts including IC4329A, the hard PL component identified in the present work is likely to have been recognized either as a partially absorbed part of the primary continuum, or relativistic reflection. After \citet{Noda2011a}, let us examine the validity of these alternative interpretations. The partially absorbed PL interpretation can be expressed as {\tt wabs[1]*powerlaw[0$^{\prime}$]}, which has the same photon index as the primary continuum {\tt powerlaw[0]}. Figure \ref{fig:c3po}d shows the simultaneous fit in which the {\tt wabs[1]*powerlaw[1]} term
comprising {\tt model\_S} is replaced with this modeling. The fit $\chi ^2 / {\rm d.\, o.\, f.\, }$ became worse from 228.4/193 to 250.5/194, yielding $\Gamma=2.08\pm0.06$ and $N_{\rm H}=(8.3\pm1.7)\times10^{23}$~cm$^{-2}$. Significant residuals are seen in the 5--6~keV continuum of the stationary part, because the PL contributing to {\tt model\_S} was required to be softer, and hence more absorbed not to over predict soft X-ray signals. Furthermore, there is even more serious problem with this alternative view; the partially absorbed PL is considered to show the same variability as the primary PL, because their sources are identical. However, the former appears in the variable part, and the latter in the stationary part. Therefore, the interpretation is self-inconsistent. This point is best understood by comparing Obs.07a and Obs.07b in figure \ref{fig:par}a; while the soft-PL in Obs.07e is $0.62\pm0.02$ times that in Obs.07a, the hard-PL flux {\it increased} marginally by a factor of $1.4^{+0.4}_{-0.3}$. In terms of the partial absorption, this would require a fine tuning between the continuum intensity and the absorption. From these results of spectroscopy and time variability, the partial covering absorption fails as a candidate for the additional component. \par
To examine yet another case of the relativistic reflection, we replaced {\tt wabs[1]*powerlaw[1]} by {\tt kdblur*pexmon[0$^{\prime}$]} where {\tt kdblur} is a convolution model (\cite{Laor1991}) to emulate relativistic blurring, and {\tt pexmon[0$^{\prime}$]} has the same shape as {\tt pexmon[0]}. In {\tt kdblur}, power-law index $\alpha$ of the radial emissivity profile and the inner disk radius $R_{\rm in}$ were allowed to vary, and the outer disk radius was fixed at $100R_{\rm g}$ where $R_{\rm g}$ is the gravitational radius. As shown in figure \ref{fig:c3po}e, the fit again became worse to $\chi ^2 / {\rm d.\, o.\, f.\, }=245.2/194$; it gave $\Gamma=2.01^{+0.03}_{-0.04}$, $R_{\rm in}\sim13R_{\rm g}$, and $\alpha=-2.1$. In particular, the relativistic reflection was unable to reproduce the sharp edge at $\sim7.0$~keV in the stationary spectrum. The value of $R_{\rm in}\sim13R_{\rm g}$is thought to be obtained as a compromise between two opposite requirements by the data; to reproduce the sharp Fe-K edge, and not to over predict narrow Fe-K lines. From these examinations, the additional component contributing to {\tt model\_S} cannot be considered to be such secondary emission, but another primary emission that is different from the $\Gamma\sim2.0$ PL. \par

\subsection{Reality of the absorption of the hard PL}
The new primary radiation as identified above carries $\sim30$\% of the total 3--10~keV flux, and stayed relatively constant (figure \ref{fig:par}) among the 2007 observations, as well as between 2007 and 2012, even though the soft PL varied significantly meanwhile. This component is harder with $\Gamma\sim1.4$, dominant in $>15$~keV, and strongly absorbed with $N_{\rm H}\sim4\times10^{23}$~cm$^{-2}$. 
Actually, as shown in figure \ref{fig:contour}a in the form of confidence contours, both $\Gamma$ and $N_{\rm H}$ are well determined typically within $\pm20$\%. \par
Then, which part of the spectrum requires such strong absorption of the hard PL? Obviously, two candidates can be considered; one is the Fe-K edge in the C3PO-derived stationary spectrum, while the other is its low-energy flux decrease. We examined these two cases, keeping the abundance of the reflection component fixed at 1 solar. To examine the former possibility, we masked the 6.8--7.2~keV band of the stationary part and repeated the simultaneous fitting. As a result, the absorption column density became from $N_{\rm H}=(4.61^{+1.41}_{-1.17})\times10^{23}$~cm$^2$ to $(4.59^{+1.43}_{-1.19})\times10^{23}$~cm$^2$, with insignificant changes in the fit goodness. Next, the $<5$~keV range of the stationary part was masked, while the Fe-K energy range was restored. This attempt yielded $N_{\rm H}=(5.14^{+1.61}_{-1.36})\times10^{23}$~cm$^2$, again with little changes in the fit goodness. Thus, the two values of $N_{\rm H}$, implied by the Fe-K edge and the low-energy bending, are consistent with each other within respective errors. Therefore, we conclude that the strong absorption on the hard PL is a robust effect, and is contributed to similar extents by the two spectral features. This agreement in $N_{\rm H}$ gives an {\it a posteriori} support to the absence of stationary signals in the 2--3~keV reference band, because this property requires the stationary spectrum to decrease below $\sim5$~keV. \par

\subsection{Comparison with other Type I Seyferts}\label{sec:others}
The additional hard primary component, thus confirmed in IC4329A and characterized as above, has actually been detected in Suzaku spectra of other AGNs as well (Noda et al. \yearcite{Noda2011a}, \yearcite{Noda2013a}, \yearcite{Noda2014}). 
In MCG--6--30--15 (\cite{Noda2011a}), for example, the 2.5--55~keV spectrum consists of a variable $\Gamma=2.1$ PL, its reflection, and another primary component with $\Gamma=1.3$. This harder primary showed low variability, and was not strongly absorbed. 
In the 2--45~keV spectrum of NGC 3516 (\cite{Noda2013a}) obtained in 2005, a variable $\Gamma=2.2$ PL, and a stationary $\Gamma=1.1$ PL absorbed with $N_{\rm H}=7.8\times10^{-22}$~cm$^{-2}$, were identified together with distant reflection. 
Yet another example is NGC 3227 (\cite{Noda2014}), in which a variable $\Gamma\sim2.3$ PL and a stationary $\Gamma\sim1.6$ PL (with $N_{\rm H}=1$--$2\times10^{-22}$~cm$^{-2}$) were found to co-exist when the source was bright and highly variable. 
Thus, the present results of IC4329A agree with these previous Suzaku achievements on several other Seyfert galaxies, and consistently reveal the 3-component nature of these objects; a soft PL with fast variability and low absorption, a harder PL with lower variability and sometimes high absorption, and a reflection component. The former two components can be regarded as primary X-ray emissions. \par
\citet{Noda2014} and \citet{NodaDron} discovered that the relative dominance of the two PL continua depends on the Eddington ratio $\eta$ of a source: 
when $\eta\lesssim0.001$, the hard PL dominates, and carries the source variability which is relatively slow and mild. As a source brightness to $\eta\gtrsim0.01$, the soft PL develops, and produces high variability on short (1--10~ks) time scales. In this respect, the present object, IC4329A, has $\eta=0.2$ (\cite{Markowitz2009}) which is higher than any other AGNs studied in Noda et al. (\yearcite{Noda2011a}, \yearcite{Noda2013a}, \yearcite{Noda2014}) and \citet{NodaDron}. Nevertheless, the hard PL was still present in this object with a 3--10~keV intensity about one third of that of the soft PL, while the fast source variability was almost carried solely by the soft PL. \par
\citet{Noda2014} argued that the softer and harder primary components are likely to arise from different regions around the SMBH: the source of the former must be closely related to the standard accretion disk, and the latter originates from a theoretically-predicted region of radiatively inefficient accretion flow (RIAF), possibly located within the truncation radius of the accretion disk. 
If so, the present results imply that the RIAF region, responsible for the hard PL production, does not disappear even when a source becomes very luminous. Further details of the suggested geometrical picture is available in figure 12 of \citet{Noda2014}.\par
The presence of the softer and harder primary components has been confirmed not only in AGNs (subsection \ref{sec:others}), but also in some Black Hole Binaries (BHBs). 
For example, \citet{Makishima2008}, \citet{Skipper2013}, and \citet{Yamada2013} revealed that the BHB Cygnus X-1 in the Low/Hard state emits softer and harder Compton components, which are considered to arise from distinct emission regions. These results of AGNs and BHBs consistently lead to a multi-zone Comptonization view, that their hard X-ray continua are emitted from multiple regions of Comptonization, and hence cannot be described by a single PL. 

\subsection{The nature of the reflection}
\begin{figure}[h!]
  \begin{minipage}{0.49\hsize}
    \begin{center}
      \includegraphics[width=8cm]{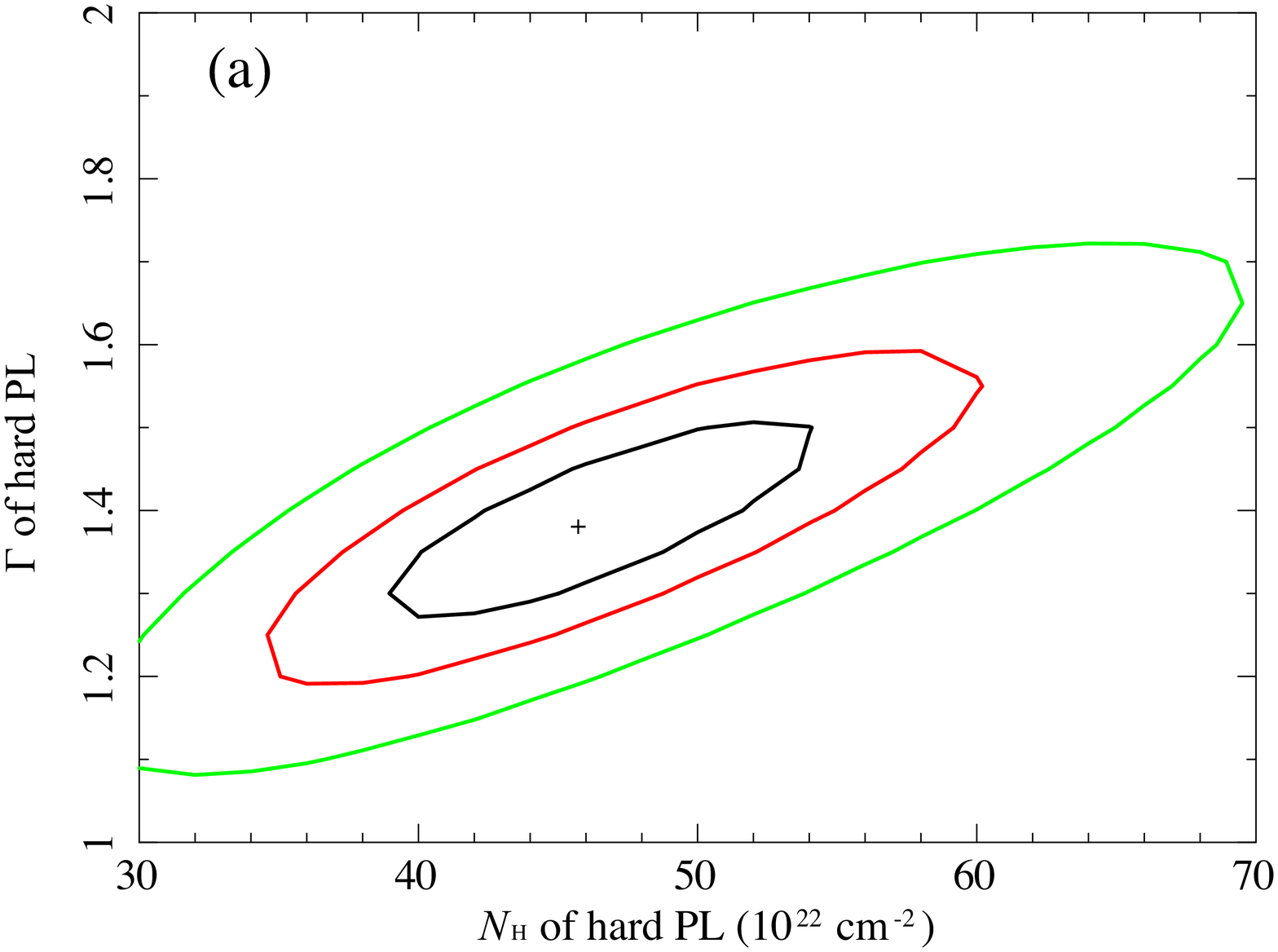}
    \end{center}
  \end{minipage}
  \begin{minipage}{0.49\hsize}
    \begin{center}
      \includegraphics[width=8cm]{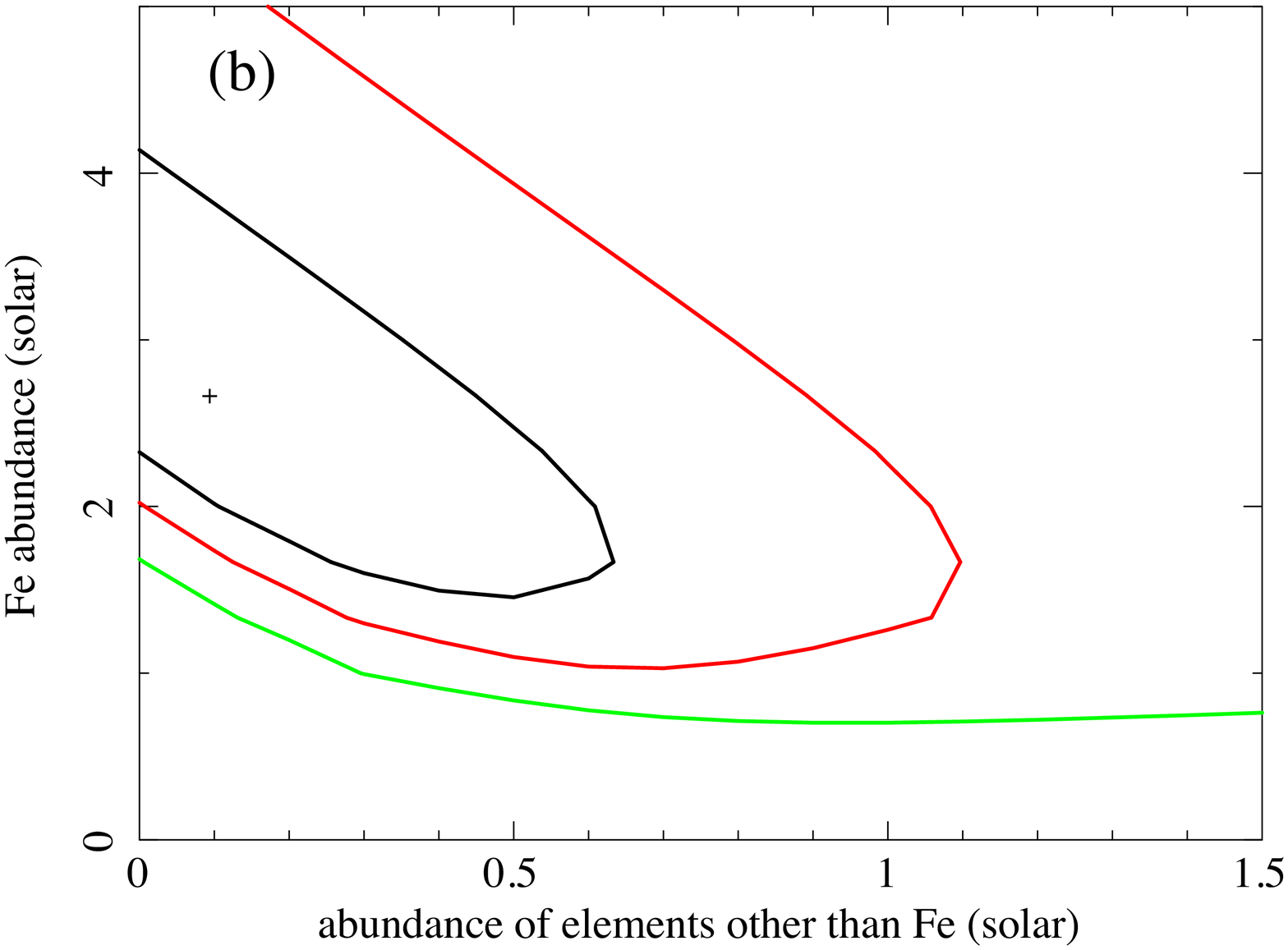}
    \end{center}
  \end{minipage}
 \caption{Confidence contours between some pairs of the spectral parameters obtained in 2007 (table \ref{tab:par}), with confidence levels of 68\% (black), 90\% (red) and 99\% (green). The reflection in this case is assumed to be produced only by the soft PL. (a) That between $N_{\rm H}$ and $\Gamma$ of the hard PL. (b) That between the abundance of metals other than Fe and the Fe abundance of the reflector, in units of solar abundances.}\label{fig:contour}
\end{figure}

While the Suzaku data do not require relativistic reflection (subsection \ref{sec:others}), the distant non-relativistic reflection, with the narrow Fe-K line, is clearly present in the data. So, let us examine whether the derived reflection parameters are reasonable or not. When the abundances of Fe and other metals (mainly C, O, Ne, Mg, Si, S and Ar with solar ratios) were fixed at 1 solar, the solid angle of the reflector was $\sim1.2\pi$, which is reasonable. When the abundances of the reflector were allowed to vary, the fit $\chi ^2 / {\rm d.\, o.\, f.\, }$ with the C3PO method decreased slightly from 228.4/193 to 223.5/191, yielding the Fe abundance of $2.7^{+3.2}_{-1.7}$ solar and that of the other metals of $0.10^{+0.98}_{-0.10}$ solar. In fact, the two abundances are consistent with 1 solar at 99\% confidence level, as shown in the confidence contour plot of figure \ref{fig:contour}b, but the best-fit values are rather derived from our expectation. \par
So far, we considered the reflection produced by the soft PL, but not that from the hard PL. We have tried ``two PLs plus two reflection" model by adding another reflection component which originates from the hard PL. 
As a result, the fit $\chi ^2 / {\rm d.\, o.\, f.\, }$ became much better from 228.4/193 to 207.3/191 and the solid angle was reduced to $0.8\pi$. The revised Fe abundance is $1.1^{+0.4}_{-0.3}$ solar and that of the others is $0.32^{+0.08}_{-0.07}$ solar, which are more reasonable. We hence conclude that the ``two PLs plus two reflection" view is more favored than the ``two PLs plus one reflection" scenario. The hard PL, with its very hard spectrum and its ability to produce reflection, is considered to come from a region near the SMBH, although further discussion on its exact location, the origin of its high absorption and its lack of fast variations are beyond the scope of the present paper. \par
These new results will provide a valuable guideline for future studies of AGNs and BHBs, using in particular the innovative X-ray observatory ASTRO-H scheduled for launch in 2016 February. \par

\section*{Acknowledgment}
This work was supported by Grant-in-Aid for JSPS Fellows Number 15J08242. 
%The hard PL also generates the reflection while it is less-variable and the harder $\Gamma$ suggests that photons with higher energies come from a region of deep gravitational potential. Therefore, the origin of the hard PL is considered to be located near the SMBH as well as that of the soft PL. 

\appendix
\section*{CCPs with other models}\label{sec:pivot}
In subsection \ref{sec:c3po}, the CCP linearity was used to conclude that the source variability is carried by intensity changes of a single component of a fixed shape (i.e., the $\Gamma=2$ PL). To better understand the implication of the CCP linearity, a simple ``Gedanken experiment" may be conducted. Let us assume that the variations are caused by {\em slope} changes of a single PL component that has a pivot at, say, $E_{\rm pv}=40$~keV. The spectral form of the variable component then becomes $A_0(E/E_{\rm pv})^{\Gamma_0+\delta\Gamma}$, where $A_0$ and $\Gamma$ are constants, whereas $\delta\Gamma$ is a time-dependent variable. By allowing the presence of an energy dependent constant (non-varying) component that is assumed as before to vanishes in the 2--3~keV reference band, this assumption has the same degrees of freedom as the case employed in subsection \ref{sec:c3po}. \par
\begin{figure}[h!]
  \begin{center}
    \includegraphics[width=8cm]{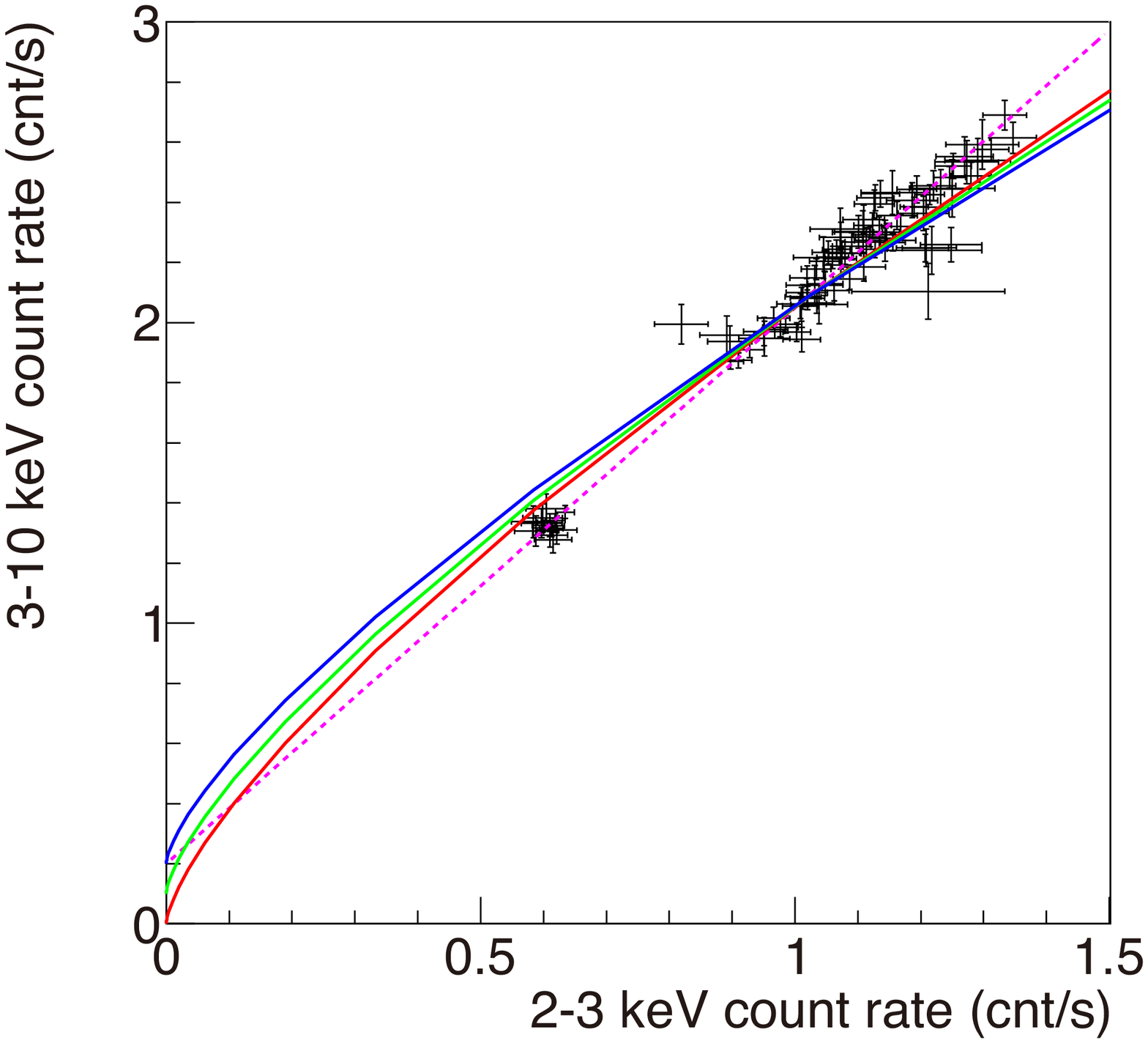}
  \end{center}
 \caption{The same CCP as figure \ref{fig:ccp}a, but compared with predictions by an alternative assumption that a single PL is varying in its slope ($\Gamma$), with a pivot at 40~keV. The three curves assume different offsets. The dotted straight line reproduces the same linear fit as in figure \ref{fig:ccp}a. }\label{fig:simccp}
\end{figure}

Referring to the 2--3~keV vs. 3--10~keV CCP in 2007 (figure \ref{fig:ccp}a), we adjusted the values of $A_0$ and $\Gamma_0$, so as to reproduce the average intensities in these two bands; this yielded $\Gamma_0=2.0$. Then, $\delta\Gamma$ was varied to mimic the source variation. The predicted locus of the source on the CCP is shown in figure \ref{fig:simccp}, where the data points are identical to those in figure \ref{fig:ccp}a. Thus, the prediction is clearly curved, and disagrees with the data. Since the predicted locus is flatter than the data distribution, allowing a positive offset in the 3--10~keV band would not help. Of course, the model locus would be brought into a better agreement with the data if $E_{\rm pv}$ is increased, but such a model assumption essentially converges to the case of the intensity variation of a constant-slope PL. Thus, the data disfavor this alternative modeling invoking a pivot and slope changes. \par

%\section{Conclusion}
%The combination of the two methods concludes that the X-ray spectra of IC4329A were well-reproduced by the model that consists of the variable PL component with $\Gamma \sim 2.0$, its reflection without high variability, and the stationary harder and strongly-absorbed PL component with $\Gamma\sim1.4$. By the fit to the time-averaged spectrum with the difference spectrum method,  we confirmed that the variable component was dominant in the reference band without any assumption. Thus, we applied C3PO method to decompose the spectrum of IC4329A to the variable and stationary parts. In a case of other targets, you can verify it with the difference spectrum method. 
% {\tt wabs[0]*(powerlaw[0]+pexmon[0]+wabs[1]*powerlaw[1]}

%%%%%%%%%%%%%%%%%%%%%%%%%%%%%%%%%%%%%%%
%\begin{table}
%  \caption{This is the first tabular.}\label{tab:first}
%  \begin{center}
%    \begin{tabular}{llll}
%      \hline
%      a & b & c & d \\
%      e & f & g & h \\
%      ....\\
%      \hline
%    \end{tabular}
%  \end{center}
%\end{table}

%\begin{longtable}{lll}
%  \caption{Sample of ``longtable"}\label{tab:LTsample}
%  \hline              
%  name & value1 & value2 \\ 
%\endfirsthead
%  \hline
%  name & value & value2  \\
%\endhead
%  \hline
%\endfoot
%  \hline
%\endlastfoot
%  \hline
%  aaaaa & bbbbb & ccccc \\
%  ...... & ..... & ..... \\
%  ...... & ..... & ..... \\
%  ...... & ..... & ..... \\
%  xxxxx & yyyyy & zzzzz \\
%\end{longtable}
%%%%%%%%%%%%%%%%%%%%%%%%%%%%%%%%%%%%%%%

%%%
% See the manual for the detail.
%%%


\begin{thebibliography}{}
% Journals(e.g. A\&A,ApJ,AJ,NMRAS,PASP ...)
% Authors, Year, Journal, Vol#, Page#
% Journal Title Abbreviation >> http://www.asj.or.jp/pasj/Jabb.html
\bibitem[Boldt (1987)]{Boldt1987}
  Boldt, E.\ 1987, Observational Cosmology, 124, 611 
\bibitem[Brenneman et al. (2014)]{Brenneman2014}
  Brenneman, L.~W., Madejski, G., Fuerst, F., et al.\ 2014, \apj, 788, 61 
% Books
\bibitem[Cerruti et al. (2011)]{Cerruti2011}
  Cerruti, M., Ponti, G., Boisson, C., et al.\ 2011, \aap, 535, A113 
\bibitem[Curazov et al. (2001)]{Churazov2001}
  Churazov, E., Gilfanov, M., \& Revnivtsev, M.\ 2001, \mnras, 321, 759 
\bibitem[George \& Fabian (1991)]{GeorgeFabian1991}
George, I.~M., \& Fabian, A.~C.\ 1991, \mnras, 249, 352
\bibitem[Fabian \& Miniutti (2005)]{FabianMiniutti2005}
  Fabian, A.~C., \& Miniutti, G.\ 2005, arXiv:astro-ph/0507409
% Books
\bibitem[Fukazawa et al. (2009)]{Fukazawa2009}
  Fukazawa, Y., Mizuno, T., Watanabe, S., et al.\ 2009, \pasj, 61, 17 
\bibitem[Gondoin et al. (2001)]{Gondoin2001}
  Gondoin, P., Barr, P., Lumb, D., et al.\ 2001, \aap, 378, 806 
\bibitem[Haardt et al. (1994)]{Haardt1994}
  Haardt, F., Maraschi, L., \& Ghisellini, G.\ 1994, \apjl, 432, L95 
\bibitem[Holt et al. (1980)]{Holt1980}
  Holt, S.~S., Mushotzky, R.~F., Boldt, E.~A., et al.\ 1980, \apjl, 241, L13 
\bibitem[Ishisaki et al. (2007)]{Ishisaki2007}
  Ishisaki, Y., Maeda, Y., Fujimoto, R., et al.\ 2007, \pasj, 59, 113 
\bibitem[Laor (1991)]{Laor1991}
  Laor, A.\ 1991, \apj, 376, 90 
\bibitem[Makishima et al.(2008)]{Makishima2008}
  Makishima, K., Takahashi, H., Yamada, S., et al.\ 2008, \pasj, 60, 585 
\bibitem[Markowitz (2009)]{Markowitz2009}
  Markowitz, A.\ 2009, \apj, 698, 1740
\bibitem[Noda (2013)]{NodaDron}
  Noda, H., 2013 PhD thesis, The University of Tokyo
\bibitem[Noda et al. (2013a)]{Noda2013a}
  Noda, H., Makishima, K., Nakazawa, K., \& Yamada, S.\ 2013a, \apj, 771, 100
\bibitem[Noda et al. (2013b)]{Noda2013b}
  Noda, H., Makishima, K., Nakazawa, K., et al.\ 2013b, \pasj, 65, 4 
\bibitem[Noda et al. (2011a)]{Noda2011a}
  Noda, H., Makishima, K., Uehara, Y., Yamada, S., \& Nakazawa, K.\ 2011a, \pasj, 63, 449 
\bibitem[Noda et al. (2011b)]{Noda2011b}
  Noda, H., Makishima, K., Yamada, S., et al.\ 2011b, \pasj, 63, 925 
\bibitem[Noda et al. (2014)]{Noda2014}
  Noda, H., Makishima, K., Yamada, S., et al.\ 2014, \apj, 794, 2
\bibitem[Perola et al. (1999)]{Perola1999}
  Perola, G.~C., Matt, G., Cappi, M., et al.\ 1999, \aap, 351, 937
\bibitem[Piro et al. (1990)]{Piro1990}
  Piro, L., Yamauchi, M., \& Matsuoka, M.\ 1990, \apjl, 360, L35
\bibitem[Skipper et al.(2013)]{Skipper2013}
  Skipper, C.~J., McHardy, I.~M., \& Maccarone, T.~J.\ 2013, \mnras, 434, 574 
\bibitem[Taylor et al. (2003)]{Taylor2003}
  Taylor, R.~D., Uttley, P., \& McHardy, I.~M.\ 2003, \mnras, 342, L31 
\bibitem[Willmer et al. (1991)]{Willmer1991}
  Willmer, C.~N.~A., Focardi, P., Chan, R., Pellegrini, P.~S., \& da Costa, N.~L.\ 1991, \aj, 101, 57 
\bibitem[Yamada et al.(2013)]{Yamada2013}
  Yamada, S., Makishima, K., Done, C., et al.\ 2013, \pasj, 65, 80 

\end{thebibliography}
\end{document}